 \RequirePackage[l2tabu, orthodox]{nag}
 \documentclass[pre,floatfix,longbibliography,superscriptaddress,twocolumn,final,notitlepage,nofootinbib]{revtex4-1}
 
 \usepackage[T1]{fontenc}
 
 \usepackage[charter]{mathdesign}
 \usepackage{amsmath}
 \usepackage[scaled=0.92]{PTSans}

 \usepackage[usenames,dvipsnames]{xcolor}
 
 \usepackage{graphicx}
 \usepackage{tabularx}
 \usepackage[caption=false]{subfig}
 \graphicspath{{./figures/}} 
 \usepackage{comment}
 \usepackage{lineno}
 \usepackage[explicit]{titlesec}
 \usepackage{footmisc}
 
 \usepackage{booktabs}
 \usepackage{multirow}
 
 \usepackage[algoruled]{algorithm2e}
 \usepackage{fancyvrb}
 \fvset{fontsize=\normalsize}
 
 \usepackage{natbib}
 
 \usepackage[colorlinks,linktoc=all,hypertexnames=false]{hyperref}

 \usepackage[capitalize]{cleveref}

 \usepackage{tabularx}
 \usepackage{graphicx}
 \usepackage{fixmath}
 \usepackage{comment}
 \usepackage{lineno}
 \usepackage[explicit]{titlesec}
 
 \usepackage{csvsimple}
 \usepackage{filecontents}

 \hypersetup{
 	bookmarks=true, 
 	unicode=false, 
 	pdftoolbar=true, 
 	pdfmenubar=true, 
 	pdffitwindow=false, 
 	pdfstartview={FitH}, 
 	pdftitle={}, 
 	pdfauthor={}, 
 	pdfcreator={}, 
 	pdfproducer={}, 
 	pdfkeywords={} {} {}, 
 	pdfnewwindow=true, 
 	colorlinks=true, 
 	citecolor=ForestGreen, 
 	filecolor=magenta, 
 	urlcolor=cyan 
 }

\newcommand{\be}{\begin{equation}}
	\newcommand{\ee}{\end{equation}} 
\newcommand{\bea}{\begin{eqnarray}}
	\newcommand{\eea}{\end{eqnarray}}

\newcommand{\e}{\mathrm{e}}

\renewcommand{\v}[1]{\mathbf{#1}} 

\newcommand{\f}[2]{\frac{#1}{#2}}

\newcommand{\bup}[1]{\left(#1\right)}
\newcommand{\rup}[1]{\left[#1\right]}

\newcommand{\id}{\mathbb{I}}

\newcommand{\ones}{\mathbf{1}}

\renewcommand{\ref}[1]{[\ref{#1}]}

\newcommand{\Htotal}{H_{\mathrm{total}}}
\newcommand{\Hself}{H_{\mathrm{self}}}
\newcommand{\kself}{k}

\usepackage[normalem]{ulem} 
\usepackage[export]{adjustbox} 
\usepackage{algcompatible} 
\usepackage{placeins}
\usepackage[export]{adjustbox}
\usepackage{soul}
\usepackage{xspace}
\usepackage{color, colortbl}
\crefname{equation}{Eq.}{Eqs.}
\crefname{section}{Sec.}{Secs.}
\crefname{figure}{Fig.}{Figs.}

\newcommand{\beginsupplement}{%
        \setcounter{table}{0}
        \renewcommand{\thetable}{A\arabic{table}}%
        \setcounter{figure}{0}
        \renewcommand{\thefigure}{A\arabic{figure}}%
        \setcounter{equation}{0}
        \renewcommand{\theequation}{A\arabic{equation}}
         \setcounter{section}{0}
        \renewcommand{\thesection}{A\arabic{section}}
 }

\newcommand{\dsrfull}{\mbox{Dynamical SpringRank}}
\newcommand{\dsr}{\textsc{DSR}\xspace}
\newcommand{\nmdsrfull}{\mbox{Offline Dynamical SpringRank}}
\newcommand{\nmdsr}{\mbox{\textsc{OffDSR}}\xspace}
\newcommand{\mwsrfull}{\mbox{moving-window SpringRank}}
\newcommand{\mwsr}{\mbox{mwSR}}

\newcommand{\opt}{{\mathrm{opt}}}
\newcommand{\Poi}{{\mathrm{Poi}}}

\begin{document}

\title{A model for efficient dynamical ranking in networks}

\author{Andrea Della Vecchia}
\email{andrea.dellavecchia@iit.it}\thanks{Contributed Equally}
\affiliation{Istituto Italiano di Tecnologia, Genoa, Italy}
\affiliation{MaLGa Center, Università di Genova, Genoa, Italy}

\author{Kibidi Neocosmos}
\email{kibidi.neocosmos@tuebingen.mpg.de}\thanks{Contributed Equally}
\affiliation{Max Planck Institute for Intelligent Systems, T\"{u}bingen, Germany}
\affiliation{African Institute for Mathematical Sciences, Cape Town, South Africa}
\affiliation{University of T\"{u}bingen, T\"{u}bingen, Germany}

\author{Daniel B. Larremore}
\email{daniel.larremore@colorado.edu}
\affiliation{Department of Computer Science, University of Colorado Boulder, Boulder, Colorado, USA}
\affiliation{BioFrontiers Institute, University of Colorado Boulder, Boulder, Colorado, USA}

\author{Cristopher Moore}
\email{moore@santafe.edu}
\affiliation{Santa Fe Institute, Santa Fe, New Mexico, USA}

\author{Caterina De Bacco}
\email{caterina.debacco@tuebingen.mpg.de}
\affiliation{Max Planck Institute for Intelligent Systems, T\"{u}bingen, Germany}


\begin{abstract}
	We present a physics-inspired method for inferring dynamic rankings in directed temporal networks---networks in which each directed and timestamped edge reflects the outcome and timing of a pairwise interaction. The inferred ranking of each node is real-valued and varies in time as each new edge, encoding an outcome like a win or loss, raises or lowers the node's estimated strength or prestige, as is often observed in real scenarios including sequences of games, tournaments, or interactions in animal hierarchies. Our method works by solving a linear system of equations and requires only one parameter to be tuned. As a result, the corresponding algorithm is scalable and efficient. We test our method by evaluating its ability to predict interactions (edges' existence) and their outcomes (edges' directions) in a variety of applications, including both synthetic and real data. Our analysis shows that in many cases our method's performance is better than existing methods for predicting dynamic rankings and interaction outcomes.
\end{abstract}
\maketitle

\section{Introduction}\label{sec:intro}

When considering a collection of people, animals, teams, or other entities, there is often an underlying hierarchy structuring the system. This hierarchy may be formally instilled in the sense that some individuals are explicitly granted certain ranks based on positions of authority. For example, in a school, there are students, teachers, and the principal or head of the school, with each position explicitly known and ranked in terms of level of authority. Alternatively, a hierarchy may be implicit in the sense that the ranks are not explicitly granted or known, but instead encoded in behaviors or interactions. For example, in animal dominance hierarchies, animals may be preferentially aggressive toward those lower in rank. In both explicit and implicit cases, hierarchies can be determined by analyzing the patterns of interactions between the entities of the system.

If we wish to infer the ranks of entities in a hierarchical structure from the patterns of their interactions, we can treat ranks as either static or dynamic, and as ordinal or real-valued. 

In the static case, time is irrelevant, and we treat all the interactions at once regardless of the sequence in which they occur, as one might when ranking the teams in a sports league at the end of a seasons. 

In the dynamic case, each individual's ranking may rise or fall over time, retaining the memory of past interactions while taking new interactions into account. This can be seen in leagues such as the U.S.\ National Basketball Association (NBA) where rankings derived from recent games provide insight for predicting games in the near future, yet the rankings themselves may nevertheless change slowly over the course of a season or seasons.
We are also interested in real-valued ranks, rather than ordinal ranks, such that the size of rank difference between two entities is an interpretable and predictive quantity, regardless of whether they are adjacent or well separated in ordinal rank.

To model systems of this type we propose \dsrfull. This builds on the previously-proposed SpringRank algorithm~\cite{de2018physical} by incorporating time information, inferring a dynamic hierarchy from a dynamic network: that is, a dataset of timestamped interactions, each of which defines a directed edge $i \to j$ indicating that $i$ ``beat'' $j$ at time $t$. We make similar physically-inspired assumptions as SpringRank, modeling directed edges as springs and assuming that entities are more likely to interact if their ranks are not too far apart. We also propose a generative model for constructing directed, hierarchical networks that evolve over time.

Finally, we evaluate \dsrfull\ on a variety of synthetic and real datasets. From our findings, we conclude that it accurately and efficiently infers ranks and predicts the direction of edges in dynamic settings. Furthermore, it frequently outperforms other algorithms such as the Elo Rating System and Whole-History Rating.


\section{Related Work}\label{sec:related_work}

Estimating hidden hierarchies from pairwise interactions is a fundamental problem in a wide variety of contexts. Several models have been proposed to study \emph{static} hierarchies, i.e., scenarios where ranks do not change in time. One line of research considers spectral methods, which exploit eigenvalues and eigenvectors of certain matrices that can be built from the network structure given an input. These methods learn real-valued scores on nodes and differ in the choice of the underlying matrix considered to solve an eigenvalue problem. Prominent examples include Eigenvector Centrality~\cite{bonacich1987}, PageRank~\cite{page1999} and Rank Centrality~\cite{negahban2016rank}.  A different line of research considers ordinal rankings, where nodes are assigned an order rather than a real-valued score. Examples are Minimum Violation Rank~\cite{ali1986minimum,slater1961inconsistencies,gupte2011finding}, Ranked Stochastic Block Model ~\cite{letizia2018resolution}, SerialRank~\cite{fogel2014serialrank} and SyncRank~\cite{cucuringu2016sync}.  
Another main line of research is that of probabilistic approaches, where a main assumption is that outcomes are random variables, and they depend on real-valued scores. These are learned using techniques from statistical inference and can be used to estimate the probability of an outcome. These approaches are considered in various domains.
For instance, in economics and psychology, Random Utility Models~\cite{train2009discrete} investigate preferences for choices that are not deterministic. A relevant example is the Bradley-Terry-Luce (BTL) model~\cite{bradley1952,luce1959}. In ecology, probabilistic niche models ~\cite{williams2010probabilistic,williams2011probabilistic,jacobs2015untangling} are used to study food webs. In social networks, a variety of probabilistic approaches have been considered. They differ in their assumptions about the underlying patterns playing a role in determining the hierarchy. For instance, social status can be considered to model friendship~\cite{ball2013friendship}. A combination of hierarchy and community structure \cite{iacovissi2021interplay} can be used to learn directed interactions between individuals. Latent space models assume that each node has a position in an underlying latent space~\cite{hoff01latentspace}. Physics-inspired models draw from analogies with physical systems, for instance a system of springs as in SpringRank~\cite{de2018physical} or continuous spin systems~\cite{cantwell-moore}.

In contrast, \textit{dynamic} approaches model dynamic environments where ranks vary in time and interactions have a relevant chronological order. For instance, the Elo Rating System~\cite{elo1978rating}, commonly used for rating chess players, is one of the most popular online methods. It was later improved by the Glicko system~\cite{glickman1995glicko}, which incorporates a measure of reliability in estimating ranks to capture their uncertainty due to, for instance, a period of inactivity or lack of data. 
The Dynamic TranSync model \cite{araya2022dynamic} assumes that observations are noisy measurements of strength differences
with zero-mean noise and imposes smoothness constraints on the time-varying strengths.
Another approach is a win-loss ranking algorithm~\cite{park2005network} and its dynamic extension~\cite{motegi2012network}. A Bayesian ranking system inferring individual ranks from team-level outcomes is the so called TrueSkill algorithm~\cite{herbrich2007trueskill}, which can be seen as a generalization of the Elo system. This has been extended by TrueSkill Through Time (TTT)~\cite{dangauthier2008trueskill} which infers smooth time series of ranks. Decaying-history ratings such as~\cite{motegi2012network} act directly on the data observations, progressively forgetting old interactions. One drawback of this approach is that time decay increases the uncertainty of player ratings: players who stop playing for a while may experience huge jumps in their ratings when they start playing again. On the other hand, players who play very frequently may have the feeling that their rating is stuck. If players do not all play at the same frequency, there is no clear way to tune the decay rate~\cite{coulom2008whole}.

Finally, an additional type of dynamic method treats ranks as time-varying, but infers the ranks at each time-step by considering the totality of all observations, including those before and after any particular time step. For instance, the Whole-History Rating (WHR)~\cite{coulom2008whole}, a Bayesian approach based on the dynamic Bradley-Terry-Luce model, computes the exact maximum a posteriori estimate of ranks over the whole history of all players. 


\section{The Model}\label{sec:model}

We represent a series of interactions between $N$ individuals as a sequence of weighted directed networks with adjacency matrix $A^t$ for $t=0,1,2,\ldots,T$. For each $t$, its entry $A_{ij}^t$ is the outcome of interactions $i \rightarrow j$ suggesting that $i$ is ranked above $j$. This allows both cardinal and ordinal inputs. For instance, in team sports, $A_{ij}^t$ could be the number of points by which team $i$ beat team $j$, or we could simply set $A_{ij}^t=1$ to indicate that $i$ won and $j$ lost. We can include the case where individuals interact multiple times at time $t$ by summing the corresponding entries.

We assume that the values of $A_{ij}^t$ are influenced by a vector of real-valued ranks $\v{s}^t=(s_{1}^t,\dots, s_{N}^t)$, where $s_i^t$ is $i$'s strength or prestige at time $t$.
To model these interactions, we follow SpringRank's approach of imagining the network as a physical system~\cite{de2018physical}. Specifically, each node $i$ is embedded in $\mathbb{R}$ at position $s_i^t$, and each directed edge $i \rightarrow j$ becomes an oriented spring with a non-zero resting length and displacement $s_i^t-s_j^t$. Since we are free to rescale latent space and the energy scale, we set the spring constant and resting length to $1$. The spring corresponding to an edge $i \rightarrow j$ at time $t$ then has energy
\be\label{eqn:staticH}
H_{ij}(s_i^t,s_j^t)=\f{1}{2} \bup{s_i^t-s_j^t-1}^{2} \, .
\ee
If there were no other effects, the total energy of the system at time $t$ would then be 
\be\label{eqn:totalstaticH}
H^t(\v{s}^t) = \sum_{i,j=1}^{N} A_{ij}^t \,H_{ij}(s_i^t,s_j^t) \, .
\ee
If we determined $\v{s}^t$ by minimizing $H^t$ for each $t$ separately, we would simply be applying the static SpringRank model separately to each ``snapshot'' of the network. This would ignore all previous (and future) interactions, and ignore the hypothesis that ranks change smoothly from one time-step to the next.

\begin{figure}[h!]
	\includegraphics[width=7cm]{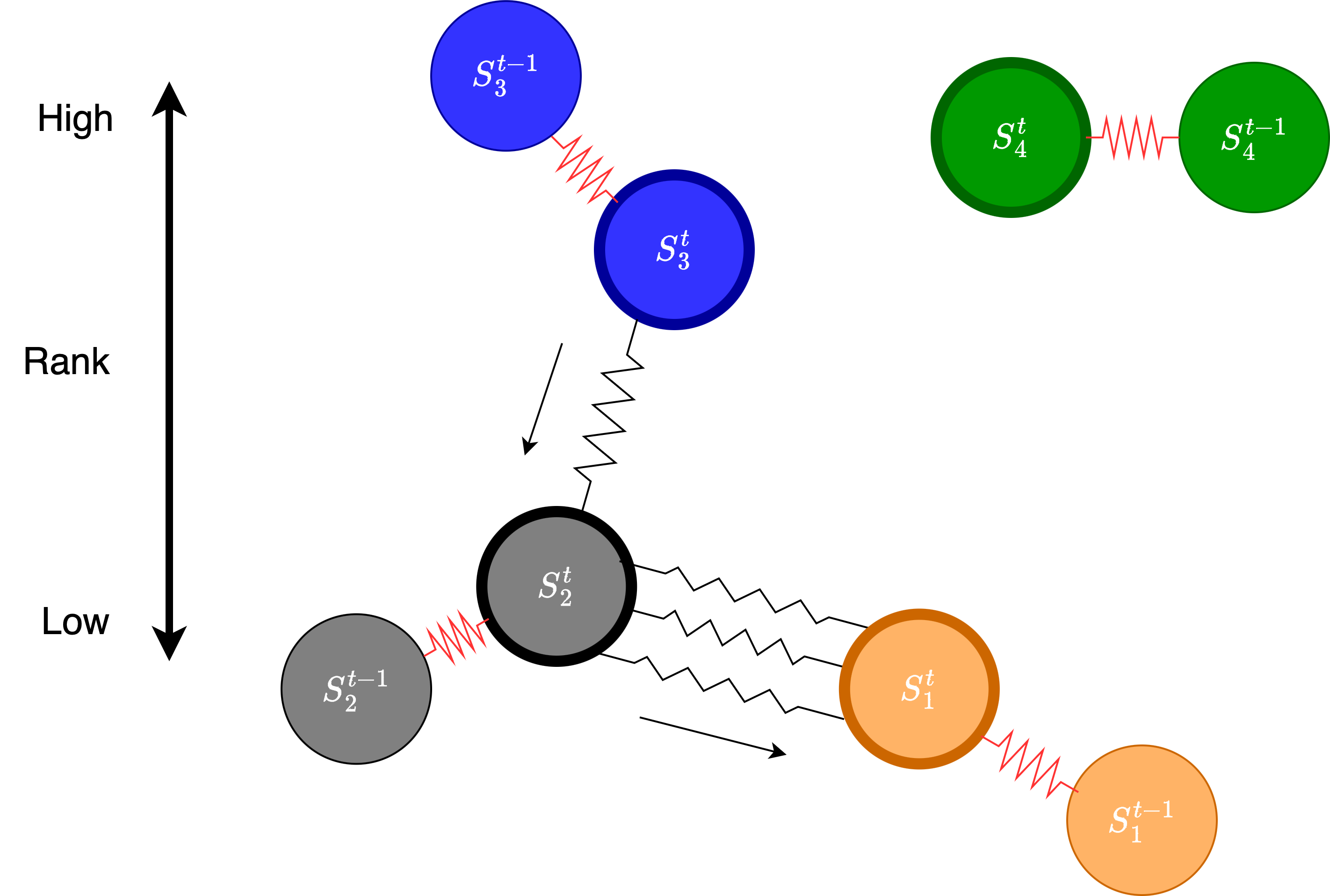}
	\caption{\textbf{A visual representation of Dynamical SpringRank.} Each node $i$ has rank $s_i$ at time $t$ and each edge is represented as a spring. The red springs indicate self-springs that connect past and present ranks. The black springs indicate interactions with different entities. The blue and grey nodes interact once while the grey and gold nodes interact three times. In contrast, the green node does not interact with the other entities. Arrows indicate the direction of a win in a directed interaction between two nodes.}
	\label{fig: dsr_visual}
\end{figure}

To model this smoothness, we also assume a dependence between ranks at successive time-steps. Specifically, we extend the Hamiltonian~\eqref{eqn:totalstaticH} with an extra term that models the \emph{self-interaction} between past and current ranks,
\begin{equation}\label{eqn:selfH}
	\Hself^t(\v{s}^t,\v{s}^{t-1}) 
	= \frac{\kself}{2} \sum_{i=1}^N (s_i^t-s_i^{t-1})^2 \, .
\end{equation}
This can be seen as a set of additional ``self-springs'' that connect the rank of each individual with its own previous rank. The spring constant $\kself$ parametrizes how smoothly we want the ranks to change from one step to the next. In inference terms, $\kself$ is a hyperparameter which we tune using cross-validation.

Summing over all time-steps $0 < t \le T$ and adding this to the pairwise interactions at each time-step then gives a total energy

\begin{align}\label{eqn:fullH}
	\Htotal(\{\v{s}^t\}) = \sum_{t=0}^T H^t(\v{s}^t) + \sum_{t=1}^T \Hself^t(\v{s}^t,\v{s}^{t-1}) \, .
\end{align}
We call this the dynamical SpringRank Hamiltonian. The optimal ranks $\v{s}^0,\v{s}^1,\ldots,\v{s}^T$ are those that minimize it.

There are two ways to minimize $\Htotal$. One is to proceed in an online way, moving forward in time. In this approach, we use the static SpringRank model Eq.~\eqref{eqn:totalstaticH} to find the initial ranks $\v{s}^0$ by minimizing $H^0(\v{s}^0)$. As in Ref.~\cite{de2018physical}, the energy is unchanged if we add a constant to all the ranks; we can break this translational symmetry by setting the mean initial rank $(1/N) \sum_{i=1}^N s_i^0$ to zero.
Then, at each subsequent time-step $t \ge 1$, we update the ranks by taking into account both the new pairwise interactions and the self-springs connecting the ranks with their previous values. Namely, given $\v{s}^{t-1}$ and $A^t$, we find the ranks $\v{s}^t$ that minimize $H^t(\v{s}^t) + \Hself^t(\v{s}^t,\v{s}^{t-1})$.

Since this is a convex function of $\v{s}^t$, we can find its minimum by setting its gradient to zero, or equivalently by balancing all the forces $s_i^t$. This yields a system of linear equations:
\begin{align}\label{eqn:fullsolution}
	\rup{ D^{out,t}+D^{in,t}- \bup{A^t + (A^t)^\dagger}+\kself\id} \,\v{s}^t
	&=\rup{D^{out,t}-D^{in,t}}\v{1} \nonumber \\& +\kself\, \v{s}^{t-1} \, . 
\end{align}

Here 
$D^{out,t}$ and $D^{in,t}$ are diagonal matrices whose entries are the weighted out- and in-degrees $D^{out,t}_{ii}=\sum_{j}A^t_{ij}$ and $D^{in,t}_{ii}=\sum_{j}A^t_{ji}$; 
$\dagger$ denotes the transpose; 
$\id$ is the identity matrix; 
and $\v{1}$ is the all-ones vector. The derivation of Eq.~\eqref{eqn:fullsolution} can be found in \Cref{sec:h_total_derivation}.

The matrix on the LHS of \cref{eqn:fullsolution} is invertible for $k>0$. This can be proved following the same reasoning as in \cite{de2018physical} under Eq.~(3) and noticing that the LHS of our \cref{eqn:fullsolution} coincides with the LHS of Eq.~(5) in \cite{de2018physical} when replacing $k$ with $\alpha$. Thus for each $A^t$ and each $\v{s}^{t-1}$, Eq.~\eqref{eqn:fullsolution} has a unique solution $\v{s}^t$. Overall, Eq.~\eqref{eqn:fullsolution} is similar to the regularized version of SpringRank~\cite{de2018physical} with regularization parameter $\alpha= \kself$. However, unlike the static model, there is a term on the right-hand side containing the previous ranks $\v{s}^{t-1}$, creating a Markovian dependence between successive time-steps. We refer to this model as \dsrfull\ (\dsr).

Importantly the online DSR approach does not actually minimize $\Htotal$, instead solving a sequence of minimization problems, one for each time step. To minimize $\Htotal$ instead, we set $\nabla \Htotal(\v{s}^t) = 0$, solving for the minimizers $\v{s}^t$ over all $N(T+1)$ ranks simultaneously, yielding the following system of equations (AI \Cref{sec:h_total_derivation_all_time}):

\begin{align}\label{eqn:h_total}
	\rup{ D^{out,t}+D^{in,t} - \bup{A^t+(A^t)^\dagger} + 2\kself\id}\,\v{s}^t 
	&=\rup{D^{out,t}-D^{in,t}}\v{1} \nonumber\\ 
	& +\kself \,\bup{\v{s}^{t-1} + \v{s}^{t+1}} \, . 
\end{align}
This differs from \Cref{eqn:fullH} in that the right-hand side now includes both past and future ranks (which doubles the contribution of $\kself$ on the left). We remove the terms $\v{s}^{t-1}$ and $\v{s}^{t+1}$ for $t=0$ and $t=T$ respectively. This is equivalent to specifying boundary conditions $\v{s}^{-1}=\v{s}^{T+1}=0$, i.e. ranks outside the considered time interval are set to zero. Other possible choices could be made for these boundary conditions. They mainly impact the values of ranks close to the boundaries and their effect lessens in the presence of many time steps. See \cref{apx:offdsr_bc} for further discussion.

This entire system has translational symmetry, since the energy Eq.~\eqref{eqn:fullH} remains the same if we add the same constant to all ranks at all times, but we can again break this symmetry by setting the mean rank to zero.

Additionally, in contrast to \Cref{eqn:fullsolution}, the ranks at $t$ now depend on both $t-1$ and $t+1$, which themselves depend on ranks at adjacent time-steps, so that ranks are affected by interactions in both the past and the future. In computer science, methods like this where the entire history is provided to the algorithm are called \emph{offline}, to distinguish them from \emph{online} approaches that update their results in real time as data becomes available. Thus we refer to this model as \nmdsrfull\ (\nmdsr).  

The cost of solving \Cref{eqn:fullsolution} for a single time-step is the same as static SpringRank with only one additional parameter to be tuned using cross-validation, and there are $T$ such $N$-dimensional equations to be solved successively. On the other hand, \Cref{eqn:h_total} requires solving a single  system of dimension $NT$, whose operator consists of $T$ blocks, each of dimension $N\times N$. While these two approaches feature numbers of non-zero entries that are fundamentally determined by the number of total edges across all time steps, the cost of solving \dsr vs \nmdsr will depend on the particular choice of linear solver~\cite{peng2021solving}.

Philosophically, Eqns.~\eqref{eqn:fullsolution} and~\eqref{eqn:h_total} are trying to do two different things. If we are given all the data $A^0,A^1,\ldots,A^T$ and we want to infer retrospectively how each individual's rank changed over time, it makes sense to include both past and future interactions as in~\eqref{eqn:h_total} so that $s_i^t$ is affected by $i$'s entire history. 
In contrast, \eqref{eqn:fullsolution} can be viewed as modeling each individual's perceived rank $s_i^t$ at a time $t \leq T$ in the past, based only on the interactions that have occurred so far, thus ignoring the future steps $t+1,\dots,T$.

In principle, one could envisage other ways to formally incorporate an explicit dependence on  $\v{s}^{t-1}$ into the model, and we provide one example in AI \Cref{sec:sidynl}. However, we found that the approaches presented in this Section provide a natural interpretation, result in good prediction performance on both real and synthetic datasets (see \Cref{sec:results}) and are computationally scalable. 

We close this section with two possible extensions to these models. First, in some settings we might have timestamps $t$ that are not successive integers $0,1,\ldots,T$. In this case, if the time interval between two successive times is $\Delta t$, one could scale the spring constant of the self-springs between time-steps as $\kself/\Delta t$. This corresponds to the fact that if we have $\Delta$ identical springs in series, each of which is stretched by $(s^t-s^{t-1})/\Delta$, their total energy is $(1/2)(\kself/\Delta)(s^t-s^{t-1})^2$. The same expression applies if the timestamps are real-valued so that $\Delta$ is not an integer.

Second, if we believe that not just the ranks themselves but their rates of change behave smoothly over time, one could add a momentum term to the Hamiltonian which is quadratic in the discrete second derivative of the ranks. Since
\begin{gather*}
	\left( (s^{t+1}-s^t) - (s^t-s^{t-1}) \right)^2
	= \left( s^{t+1} - 2 s^t + s^{t-1} \right)^2 \\
	= 2 (s^t-s^{t-1})^2 + 2 (s^{t+1}-s^t)^2 - (s^{t+1} - s^{t-1})^2 \, ,
\end{gather*}
this is equivalent to adding a repulsive force, i.e., a spring with negative spring constant, between ranks two time-steps apart. Note that the system nevertheless remains convex: this momentum term is positive semidefinite, so adding it to~\eqref{eqn:fullH} keeps the coupling matrix positive definite except for translational symmetry. Of course, these terms are second-order in time. In the online approach, one would have to determine $\v{s}^0$ from the static model, $\v{s}^1$ from the first-order model~\eqref{eqn:fullsolution}, and then use the model including this momentum term for $\v{s}^t$ for $t \ge 2$. We have not pursued this here, but it may make sense for certain datasets.

\subsection{Moving-window SpringRank}\label{subsec:mwsr}

Before we test the various versions of \dsrfull\ defined above, we consider a simpler model as a baseline. 
The simplest way to extend SpringRank to a dynamical context is to apply the static model to the interactions in a series of ``windows,'' where in each window we sum the interactions over a series of consecutive time-steps. For instance, we can compute $\v{s}^t$ for each $t$ by applying the static model to a window of width $\tau$, i.e., replacing $A^t$ with $\sum_{t'=t}^{t+\tau-1} A^{t'}$. Since these windows overlap, the resulting estimates $\v{s}^t$ will be smooth to some extent, even without imposing an explicit dependence between $\v{s}^t$ and $\v{s}^{t-1}$. We use this method, which we call \mwsrfull\ (\mwsr), as a baseline to compare with the dynamical models presented above.

Roughly speaking, a larger $\tau$ is like a larger self-spring constant $\kself$, since it induces more overlap between windows and thus a stronger correlation between the inferred ranks. However, like a decaying-history approach, \mwsr\ assumes a particular kernel for the importance of past time-steps: namely, that all $t'$ in the window are equally important. In contrast, \dsrfull\ infers the importance of past time-steps by coupling $\v{s}^t$ with $\v{s}^{t-1}$.

However, both models have a free parameter that needs to be tuned, i.e., $\kself$ and $\tau$. A shorter window $\tau$ or smaller spring constant $\kself$ allows the ranks to respond quickly to new interactions, while a longer window or larger spring constant more tightly couples nearby estimates. This trade-off suggests the existence of an optimal window length $\tau_{\opt}$. We tune $\tau$ using a cross-validation procedure as explained in AI \Cref{sisec:tuning}.

\subsection{Generative Model and Synthetic Data}
\label{sec:genmod}

Analogous to a model presented in~\cite{de2018physical}, we propose a probabilistic generative model for dynamic data. It takes as input the ranks $\v{s}^t$ and generates a sequence of weighted directed networks with adjacency matrix $A^t$ at time $t$. One can also imagine models that generate the ranks, for instance with a random walk with Gaussian steps whose log-probability is the self-spring Hamiltonian~\eqref{eqn:selfH}, but we treat $\v{s}^t$ as an input since we want the user of this model to have control over how the ground-truth ranks vary with time.  For instance, in our experiments below we generate synthetic data where the ranks vary sinusoidally.

The generative model has two real-valued parameters: a signal-to-noise ratio or inverse temperature $\beta$, and an overall density of edges $c$. Given the ranks $\v{s}^t$, it generates weighted, directed edges between each pair of nodes $i,j$ independently, as follows. The probability $P_{ij}^t(\beta)$ of $i$ ``beating'' $j$ at time $t$, giving a directed edge $i \to j$, is a logistic function as in~\cite{de2018physical} or the Bradley-Terry-Luce model~\cite{bradley1952,luce1959}:
\bea
\nonumber P_{ij}^t(\beta)=\frac{1}{1+\e^{-2\beta(s_i^t-s_j^t)}} \, .
\eea
The number of such edges, which gives the integer weight $A_{ij}^t$, is then drawn from a Poisson distribution whose mean $\lambda_{ij}^t$ is $cP^t_{ij}\,(\beta)$: 
\be
\label{generative_poiss}
A^t_{ij} \sim \Poi\left(\lambda_{ij}^t=\frac{c}{1+\e^{-2\beta(s_i^t-s_j^t)}}\right).
\ee
Since $P_{ij}^t(\beta) + P_{ji}^t(\beta)=1$, for any pair $i,j$ the total number of interactions $A_{ij}^t + A_{ji}^t$ is Poisson-distributed with mean $c$. The rank differences $s_i^t-s_j^t$ are used only to choose the directions of these edges. This  is equivalent to a model where we define a random multigraph where the number of edges between $i$ and $j$ is $\Poi(c)$, and then we choose the direction of each edge independently according to $P_{ij}^t$.

This is different from the generative model proposed in the static case in~\cite{de2018physical}. In that model the probability that $i$ and $j$ interact depends on $s_i-s_j$ so that nodes are more likely to interact if their ranks are fairly close. This is consistent with SpringRank's assumption that if $i$ beats $j$ then $j$ is below $i$, but not too far below it (since the springs have resting length $1$). This assumption makes sense for some datasets but not for others. By generating synthetic data without this dependence, our intent is to pose a greater challenge to SpringRank by modeling (for example) round-robin tournaments where every team plays each other.

\subsection{Model Evaluation}
\label{sec:testing}

Assessing a ranking model on real datasets is not straightforward since we do not know the true values of the underlying ranks. Nevertheless, we may measure the extent to which inferred ranks are accurate in the sense that they can predict the outcome of new observations. 

There are several performance metrics that can be used for prediction evaluation. From coarse-grained measures capable of predicting the likely winner to more fine-grained measures that also estimate odds, we consider four main metrics in our experiments, detailed in \Cref{sisec:evaluation}. We measure prediction performance using a cross-validation protocol where datasets are divided into training and test sets. The training set is used for hyperparameter tuning and parameter estimation while performance is evaluated on the test set. In order to preserve the chronological ordering of the data, the test set contains future observations, i.e., observations that chronologically follow those used in training. Hyperparameters for each method are tuned using grid-search in order to maximize the performance metrics as described in AI \Cref{sisec:tuning}.


\section{Results}
\label{sec:results}

Having introduced \dsrfull\ and its generative counterpart, as well as discussing model selection between the dynamic and static versions of SpringRank, we now illustrate their behavior on synthetic and real data.

We compare prediction performance on held-out test data for \dsr\ and \nmdsr\ against several state-of-the-art algorithms such as the Elo Rating System (Elo)~\cite{elo1978rating}, TrueSkill (TS)~\cite{dangauthier2008trueskill}, ``win-loss'' decay-history rating (W-L)~\cite{motegi2012network} and Whole-History Rating (WHR)~\cite{coulom2008whole} (see Sec.~\ref{apx:description_algs} for a brief description of these methods) . In addition, we consider two baselines: static SpringRank (SR) \cite{de2018physical} and \mwsr\ presented above in \Cref{subsec:mwsr}.  (Note that static SpringRank is the limiting case of \mwsr\ with one window covering the entire dataset.)
Additionally, since \nmdsr considers future information, in the experiments it was only given past information so that a fair comparison can be made with the other models in terms of prediction performance (see \Cref{apx:offdsr_bc} for a further discussion).

\subsection{Performance on Synthetic Data}
\label{sec:resultssynt}

We first consider synthetic data, generated as described in \Cref{sec:genmod}, in which ranks evolve according to periodic ground truth dynamics,
\begin{equation}
	\label{eq:dyn_score}
	s_{i}^{t}=b_{i}\cos(\omega_i t+\phi_i) + c_{i}\cos(\upsilon_i t+\phi_i)\ ,
\end{equation}
where $b_{i}$,$c_{i}$, $\omega_i$, $\phi_i$, $\upsilon_i$ are parameters randomly chosen for each node from a continuous uniform distribution (see \Cref{apx:synthetic_experiments} for details). This results in changes in rankings, and swaps in the order of ranks, reminiscent of real scenarios where teams and players rise and fall. The fact that we assign individual parameters to nodes allows us to mimic realistic scenarios where different teams change their ranks at different rates during a season. For instance, some teams can have more constant ranks while others can change more rapidly.

In order to assess the effect of different network structures, we vary parameters $\beta$ and $c$ from Eq.~\eqref{generative_poiss}. We tabulate the results in \Cref{tab:varying_noise} for varying values of $\beta$ and fixed $c=0.5$, and in \Cref{tab:varying_density} for varying values of $c$ and fixed $\beta=2.0$. We use $50\%$ of the data for training and 4 time-steps for testing, detailed in \Cref{sisec:tuning}.

Overall, \dsr\ has the largest number of top performances when considering all metrics (Tables~\ref{tab:varying_noise} and \ref{tab:varying_density}). Notably, \dsr\ outperforms its offline variant \nmdsr, even though \nmdsr\ is given the entire history. This implies that using future interactions to retrodict out-of-sample interactions is less accurate than simply using past interactions. Recall also that \dsr\ is more efficient algorithmically than \nmdsr. Overall, all algorithms perform better for higher values of $\beta$ (i.e., lower noise).

The model with the second largest number of top performances is WHR, which does well particularly for $\sigma_{L}$, the metric that accounts for the likelihood of the outcomes. Notably, static SpringRank is significantly worse than the other models, illustrating that performance can be negatively affected by choosing a static model in dynamical settings. However, for higher noise levels such as $\beta=0.1$, static SpringRank performs comparably well to the other models. This suggests that when there is less structure in the data, a static algorithm is enough: taking the chronological order of events into account does not improve performance. 

As a sanity check of our permutation test for model selection between static and dynamic models, we also considered synthetic datasets generated with static ranks $s^{t}_{i}=s_i$. 
As expected, static SpringRank performs well in comparison to the dynamic algorithms as shown in \Cref{tab:static_results}. 

\begin{figure}[t]
	\includegraphics[width=\linewidth]{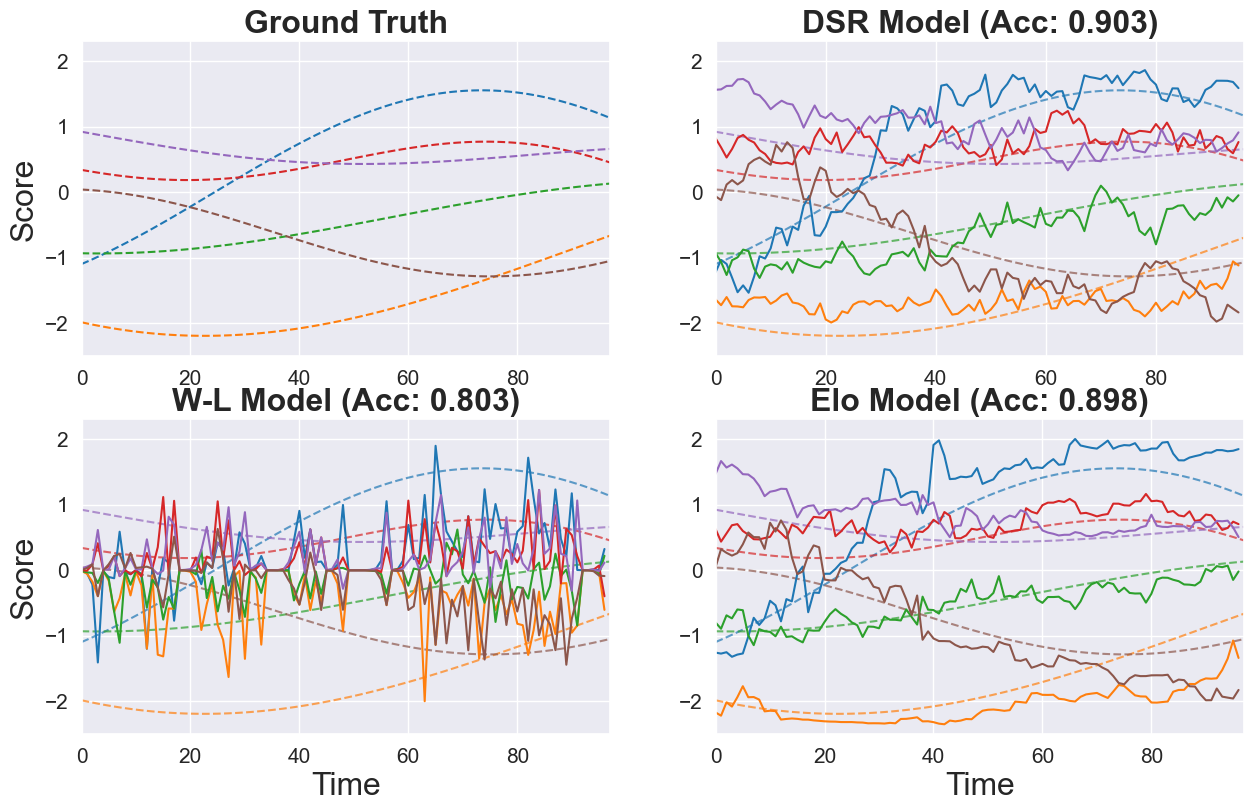}
	\caption{\textbf{Evolution of inferred ranks over time on synthetic data.} We illustrate the inferred ranks of three models over time: \dsr, W-L and Elo. We also illustrate the ground truth of the synthetic ranks over time as a comparison (top left). The synthetic data is generated by setting $\beta=2.0$ and $c=0.5$. Dashed lines are ground truth ranks.}
	\label{fig:synscores}
\end{figure}

Finally, we qualitatively investigate the inferred ranking in \Cref{fig:synscores} for \dsr, Elo and W-L where the hierarchy as well as predictive performance is the strongest, as can be seen in \Cref{tab:varying_noise} when $\beta=2.0$. We notice how the time-scale of the evolution of the ranks is different in all cases, with W-L having frequent and sudden jumps while \dsr\ and Elo are smoother with roughly equal performance. In all cases though, we notice small jumps indicating changes in ranks that deviate from the smoothness in the ground truth. Nevertheless performance is strong for \dsr\ and Elo, who perform roughly equally well, as the behaviors of the individual trajectories resembles that of the ground truth well in both cases.

These synthetic tests suggest that dynamical algorithms capture relevant information when the data has a hierarchical structure and chronological ordering matters (i.e low noise). In these cases,  \dsrfull\ performs the best according to several metrics. For higher noise levels or static ranks, timestamp information is no longer relevant and static SpringRank performs well.

\begin{table}[t]
	\resizebox{\linewidth}{!}{
		\begin{tabular}{c|c|cccccccc}
			{$\boldsymbol{\beta}$} & {\textbf{Metric}} & {\textbf{Elo}} & {\textbf{\nmdsr}} & {\textbf{\mwsr}} & {\textbf{\dsr}} & {\textbf{SR}} & {\textbf{TS}} & {\textbf{W-L}} & {\textbf{WHR}} \\
			\hline
			\multirow[t]{4}{*}{\textbf{0.1}} & accuracy & 0.545 & 0.544 & 0.533 & \sethlcolor{green}\hl{\textbf{0.549}} & 0.540 & 0.548 & 0.511 & 0.549 \\
			\rowcolor[gray]{0.95}[\tabcolsep][\tabcolsep]
			& agony & 1.568 & 1.596 & 1.658 & 1.574 & 1.646 & \sethlcolor{green}\hl{\textbf{1.551}} & 1.784 & 1.578 \\
			& $\sigma_a$ & 0.593 & \sethlcolor{green}\hl{\textbf{0.594}} & 0.576 & 0.584 & 0.594 & 0.593 & --   & 0.592 \\
			\rowcolor[gray]{0.95}[\tabcolsep][\tabcolsep]
			& $\sigma_L$ & -1.426 & -1.382 & -1.389 & \sethlcolor{green}\hl{\textbf{-1.378}} & -1.382 & -1.392 & --   & -1.389 \\
			\hline
			\multirow[t]{4}{*}{\textbf{0.5}} & accuracy & 0.700 & 0.700 & 0.698 & 0.700 & 0.652 & \sethlcolor{green}\hl{\textbf{0.703}} & 0.620 & 0.701 \\
			\rowcolor[gray]{0.95}[\tabcolsep][\tabcolsep]
			& agony & 0.881 & 0.877 & 0.887 & \sethlcolor{green}\hl{\textbf{0.877}} & 1.075 & 0.885 & 1.230 & 0.882 \\
			& $\sigma_a$ & 0.666 & 0.647 & \sethlcolor{green}\hl{\textbf{0.708}} & 0.705 & 0.635 & 0.674 & -- & 0.670 \\
			\rowcolor[gray]{0.95}[\tabcolsep][\tabcolsep]
			& $\sigma_L$ & -1.344 & -1.286 & -1.165 & -1.163 & -1.263 & -1.167 & -- & \sethlcolor{green}\hl{\textbf{-1.152}} \\
			\hline
			\multirow[t]{4}{*}{\textbf{1.0}} & accuracy & 0.810 & \sethlcolor{green}\hl{\textbf{0.816}} & 0.810 & 0.810 & 0.713 & 0.808 & 0.721 & 0.811 \\
			\rowcolor[gray]{0.95}[\tabcolsep][\tabcolsep]
			& agony & 0.455 & 0.436 & \sethlcolor{green}\hl{\textbf{0.429}} & 0.440 & 0.799 & 0.458 & 0.766 & 0.442 \\
			& $\sigma_a$ & 0.771 & 0.783 & 0.813 & \sethlcolor{green}\hl{\textbf{0.813}} & 0.702 & 0.767 & -- & 0.756 \\
			\rowcolor[gray]{0.95}[\tabcolsep][\tabcolsep]
			& $\sigma_L$ & -1.143 & -0.988 & -0.848 & -0.853 & -1.149 & -0.863 & -- & \sethlcolor{green}\hl{\textbf{-0.846}} \\
			\hline
			\multirow[t]{4}{*}{\textbf{1.5}} & accuracy & \sethlcolor{green}\hl{\textbf{0.866}} & 0.862 & 0.863 & 0.864 & 0.752 & 0.865 & 0.772 & 0.863 \\
			\rowcolor[gray]{0.95}[\tabcolsep][\tabcolsep]
			& agony & \sethlcolor{green}\hl{\textbf{0.260}} & 0.269 & 0.269 & 0.261 & 0.655 & 0.266 & 0.546 & 0.270 \\
			& $\sigma_a$ & 0.835 & 0.823 & 0.863 & \sethlcolor{green}\hl{\textbf{0.866}} & 0.745 & 0.825 & -- & 0.815 \\
			\rowcolor[gray]{0.95}[\tabcolsep][\tabcolsep]
			& $\sigma_L$ & -0.883 & -0.918 & -0.671 & -0.670 & -1.128 & -0.662 & -- & \sethlcolor{green}\hl{\textbf{-0.655}} \\
			\hline
			\multirow[t]{4}{*}{\textbf{2.0}} & accuracy & 0.898 & 0.898 & 0.898 & \sethlcolor{green}\hl{\textbf{0.903}} & 0.772 & 0.900 & 0.803 & 0.900 \\
			\rowcolor[gray]{0.95}[\tabcolsep][\tabcolsep]
			& agony & 0.172 & 0.179 & 0.171 & \sethlcolor{green}\hl{\textbf{0.163}} & 0.606 & 0.172 & 0.451 & 0.169 \\
			& $\sigma_a$ & 0.876 & 0.847 & 0.899 & \sethlcolor{green}\hl{\textbf{0.901}} & 0.769 & 0.861 & -- & 0.856 \\
			\rowcolor[gray]{0.95}[\tabcolsep][\tabcolsep]
			& $\sigma_L$ & -0.673 & -0.844 & -0.492 & -0.500 & -1.088 & -0.500 & -- & \sethlcolor{green}\hl{\textbf{-0.492}} \\
	\end{tabular}}
	\caption{\textbf{Results obtained from synthetic data with varying noise levels, represented by $\boldsymbol{\beta}$.} Each value is the mean of 4 independent realizations of the noisy model. The green highlighted values are the top performances for the considered metric. Notably, some of the values in the same row appear identical but only a single value is highlighted. The reason for this is that the highlighted value is better by less than three decimal places. \Cref{tb:sem_beta} contains the standard error of the above values. $\sigma_a$ and $\sigma_L$ cannot be applied to the W-L model, so there are no values for the metrics.}
	\label{tab:varying_noise}
\end{table}

\subsection{Performance on Real Data}
\label{sec:resultsreal}

We consider a variety of real datasets of timestamped interactions, as described in \Cref{tb:datasets}. These datasets come from competitions in well-known sports such as soccer, basketball and chess. They are both relevant and relatable sources of information for our experiments. 

In soccer, we consider the Italian Serie A and the English Premier League (EPL). The Serie A data is from the period 1993--2016 and contains the results of thousands of  games between 47 teams. Similarly, the EPL contains results of thousands of games between 39 teams in the period 2006--2018. In contrast, the NBA dataset contains roughly three times the number of  EPL matches from 2010-2018 between 30 teams. All these 3 datasets can be found on \textit{kaggle.com}. Finally, the chess dataset is obtained from matches on \textit{lichess.org}. It contains 298 matches from 2014-2017. In all cases, $A^{t}_{ij}$ is the number of times team $i$ (or for chess, player $i$) beats $j$ in a given time-step $t$. The definition of a time-step varies from sport to sport (see below).

\begin{table}[h!]
	\caption{Descriptions of the real datasets.}
	\label{tb:datasets}
	\vskip 0.15in
	\begin{center}
		\begin{small}
			\begin{tabular}{lllll}
				\toprule
				Competition	&   Type & $N_{teams}$  &  $N_{games}$ & $T_{steps}$\\
				\midrule
				NBA &       Basketball &  30 & 9594& 218 \\
				lichess.org &       Chess & 96  & 298& 90 \\
				Serie A                      &  Soccer &  47  &  5679 & 397 \\
				English Premier League (EPL) &  Soccer &  39  &  3396 & 114 \\
				\bottomrule
			\end{tabular}
		\end{small}
	\end{center}
	\vskip -0.1in
\end{table}

As with synthetic data, we found that \dsr\ outperforms the other algorithms in terms of the most top performances across our four different metrics (\Cref{tab:real_data_results}). Elo and WHR are the next best performers: Elo does slightly better in a few cases on the accuracy or agony metric for NBA and chess, and as in the synthetic data WHR does well for the $\sigma_L$ metric, the conditional log-likelihood of generating directed edges (outcomes) given their existence.

\begin{table}[h!]
	\resizebox{\linewidth}{2.5cm}{	
		\begin{tabular}{cc|cccccccccc}
			{\textbf{Dataset}} & {\textbf{Metric}} & {\textbf{Elo}}  & {\textbf{\nmdsr}} & {\textbf{\mwsr}} & {\textbf{\dsr}} & {\textbf{SR}} & {\textbf{TS}} & {\textbf{WHR}} & {\textbf{W-L}} \\
			\hline
			\multirow[t]{4}*{\textbf{NBA}}  & accuracy & \sethlcolor{green}\hl{\textbf{0.650}} & 0.642 & 0.637 & 0.649 & 0.607 & 0.645 & 0.648 & 0.565 \\
			\rowcolor[gray]{0.95}[\tabcolsep][\tabcolsep]
			& agony & \sethlcolor{green}\hl{\textbf{2.981}} & 3.050 & 3.084 & 2.987 & 3.568 & 3.006 & 2.997 & 4.071 \\
			& $\sigma_a$ & 0.579 & 0.562 & 0.639 & \sethlcolor{green}\hl{\textbf{0.646}} & 0.596 & 0.584 & 0.580 & --   \\
			\rowcolor[gray]{0.95}[\tabcolsep][\tabcolsep]
			& $\sigma_L$ & -1.426 & -1.330 & -1.266 & -1.256 & -1.324 & -1.280 & \sethlcolor{green}\hl{\textbf{-1.255}} & --   \\
			\hline
			\multirow[t]{4}{*}{\textbf{Chess}} & accuracy & \sethlcolor{green}\hl{\textbf{0.677}} & 0.633 & 0.637 & 0.665 & 0.672 & 0.665 & 0.647 & 0.539 \\
			\rowcolor[gray]{0.95}[\tabcolsep][\tabcolsep]
			& agony & 8.404 & 11.641 & 9.242 & 8.470 & 8.074 & 7.693 & 8.179 & \sethlcolor{green}\hl{\textbf{1.087}} \\
			& $\sigma_a$ & 0.615 & 0.580 & 0.626 & \sethlcolor{green}\hl{\textbf{0.651}} & 0.581 & 0.628 & 0.626 & --   \\
			\rowcolor[gray]{0.95}[\tabcolsep][\tabcolsep]
			& $\sigma_L$ & -1.290 & -1.341 & -1.294 & -1.333 & -1.550 & -1.255 & \sethlcolor{green}\hl{\textbf{-1.206}} & --   \\
			\hline
			\multirow[t]{4}{*}{\textbf{EPL}} & accuracy & 0.678 & \sethlcolor{green}\hl{\textbf{0.681}} & 0.669 & 0.675 & 0.679 & 0.672 & 0.673 & 0.609 \\
			\rowcolor[gray]{0.95}[\tabcolsep][\tabcolsep]
			& agony & 3.239 & 4.144 & 4.141 & \sethlcolor{green}\hl{\textbf{3.147}} & 3.401 & 3.797 & 3.825 & 5.438 \\
			& $\sigma_a$ & 0.595 & 0.530 & 0.669 & \sethlcolor{green}\hl{\textbf{0.675}} & 0.662 & 0.601 & 0.598 & --   \\
			\rowcolor[gray]{0.95}[\tabcolsep][\tabcolsep]
			& $\sigma_L$ & -1.285 & -1.357 & -1.208 & \sethlcolor{green}\hl{\textbf{-1.184}} & -1.206 & -1.211 & -1.199 & --   \\
			\hline
			\multirow[t]{4}{*}{\textbf{Serie A}} & accuracy & 0.655 & 0.652 & 0.630 & 0.653 & \sethlcolor{green}\hl{\textbf{0.663}} & 0.655 & 0.653 & 0.564 \\
			\rowcolor[gray]{0.95}[\tabcolsep][\tabcolsep]
			& agony & 4.296 & 5.800 & 6.278 & \sethlcolor{green}\hl{\textbf{4.041}} & 4.241 & 5.669 & 5.653 & 8.101 \\
			& $\sigma_a$ & 0.582 & 0.530 & 0.628 & \sethlcolor{green}\hl{\textbf{0.652}} & 0.647 & 0.590 & 0.585 & --   \\
			\rowcolor[gray]{0.95}[\tabcolsep][\tabcolsep]
			& $\sigma_L$  & -1.363 & -1.357 & -1.287 & -1.240 & -1.269 & -1.257 & \sethlcolor{green}\hl{\textbf{-1.237}} & --   \\
	\end{tabular}}
	\caption{\textbf{Results obtained from real data.} The green highlighted values are the top performances for the considered metric. Notably, some of the values in the same row appear identical but only a single value is highlighted. The reason for this is that the highlighted value is better by less than three decimal places. \Cref{tb:sem_real} contains the standard error of the above values.  $\sigma_a$ and $\sigma_L$ cannot be applied to the W-L model hence there are no values for the metrics.}
	\label{tab:real_data_results}	
\end{table}

Perhaps surprisingly, static SpringRank performs well on both the Serie A and chess datasets, achieving the highest accuracy on Serie A. For Serie A, this could, in part, be explained by the fact that the dataset has a lower frequency of matches compared to the NBA. In a soccer competition, typically matches are played weekly, while in the NBA teams play more frequently, two or three times per week. It could be that a lower frequency implies fewer dependencies between time-steps, thus making a dynamical model that implies a dependence between time-steps less expressive. 
At the same time, the regulations behind the European soccer leagues and the NBA are quite different (with salary caps and college drafts aiming at levelling the teams' strength in the NBA). This could imply a more constant ranking in soccer than in the NBA, making a static model work well in practice. In fact, in the last 20 years in Serie A only four teams won the title and only six in the English Premier League, sometime with long winning streaks for an individual team. On the contrary, NBA championships are clearly more unpredictable, with ten different winners in last twenty years, with a maximum of two titles won consecutively by the same team. 

For the chess dataset, each time-step represents a day of matches, but match days are not necessarily consecutive. For example, the first day of matches is 2014-03-04 and the second is 2015-11-15. Again, this poses the problem of large gaps in time which could lessen the connection between time-steps. 

As such, in both the Serie A and chess datasets, it is understandable that the static version of SpringRank would perform fairly well as time-steps do not influence each other as much as in, for example, the NBA dataset. This is further supported by the closeness in results between the static version of SpringRank and the dynamic models on the soccer and chess datasets. In contrast, the gap of results from  the NBA  dataset between the aforementioned static and dynamic models is larger. We discuss the influence of time further in \Cref{sec:dynamicity}. (The Serie A and chess datasets might also be suitable for the model described above where time intervals between snapshots can vary; we leave this for future work.)

\begin{figure}[t]
	\centering
	\includegraphics[width=9cm]{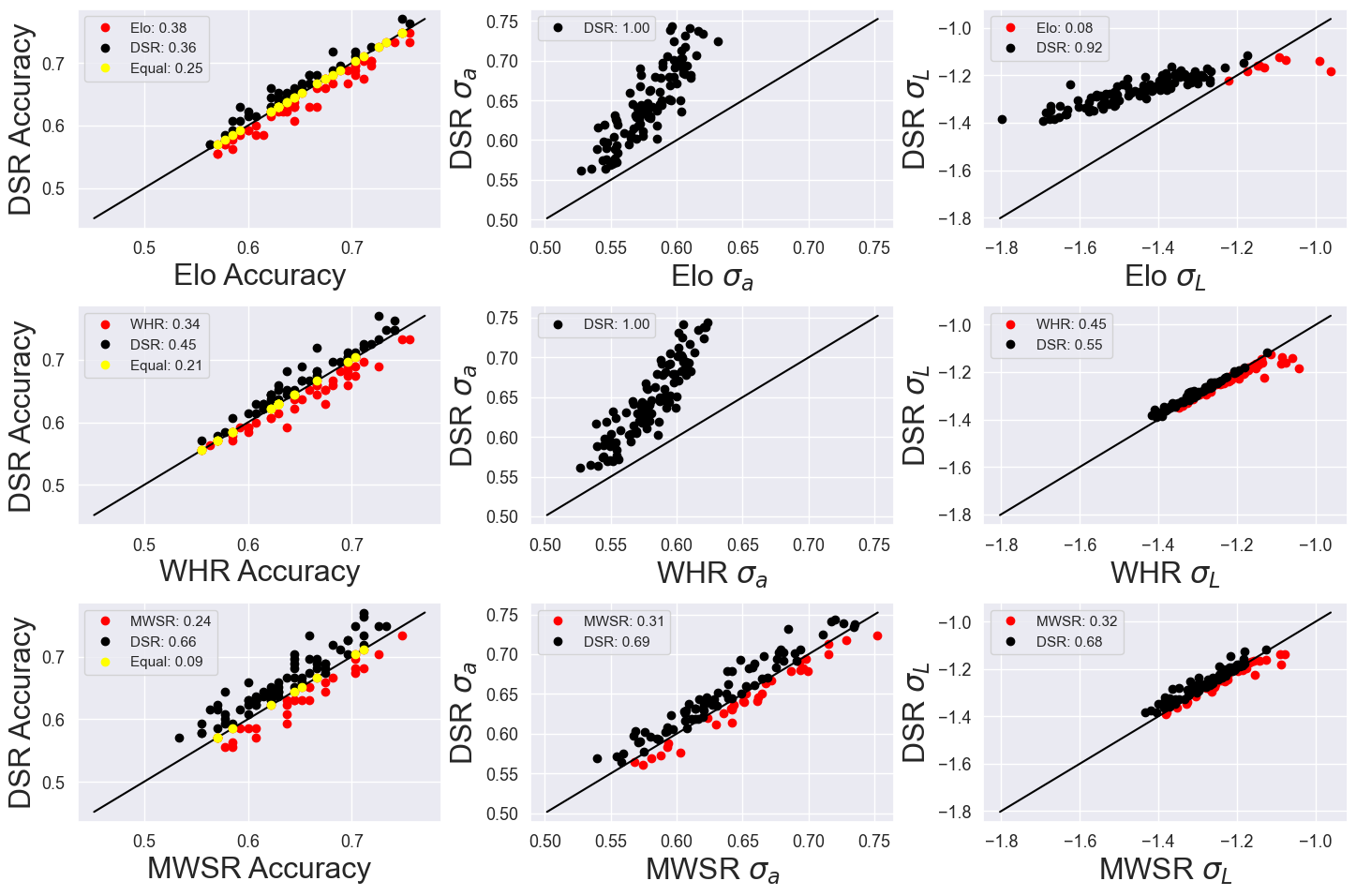}
	\caption{\textbf{Fold-by-fold evaluation on the NBA dataset.} We compare the predictions of \dsr\ to Elo, WHR and \mwsr\ in relation to the performance metrics $\sigma_a$, $\sigma_L$ and accuracy. The black points above the diagonal represent folds where \dsr\ outperformed its competitors; yellow points indicate equal performance and red points represent \dsr\ loses (where it was outperformed by competitors). Numbers inside the legend are the number of trials that an algorithm outperforms the other in percentage.}
	\label{fig:test_comparison}
\end{figure}

Overall, we observe a fairly broad distribution of values for the various metrics over the cross-validation trials, as there are matches that are more difficult to predict than others. Hence, we take a closer look by analyzing a fold-by-fold performance comparison, where we assess the number of test sets in which one algorithm outperforms the others. We find that \dsr\ performs equal to or better than the other algorithms in most cases on the NBA dataset, and in all cases when compared to Elo and WHR in terms of $\sigma_{a}$ (\Cref{fig:test_comparison}).

\begin{figure*}[t]
	\centering
	\includegraphics[width=\textwidth]{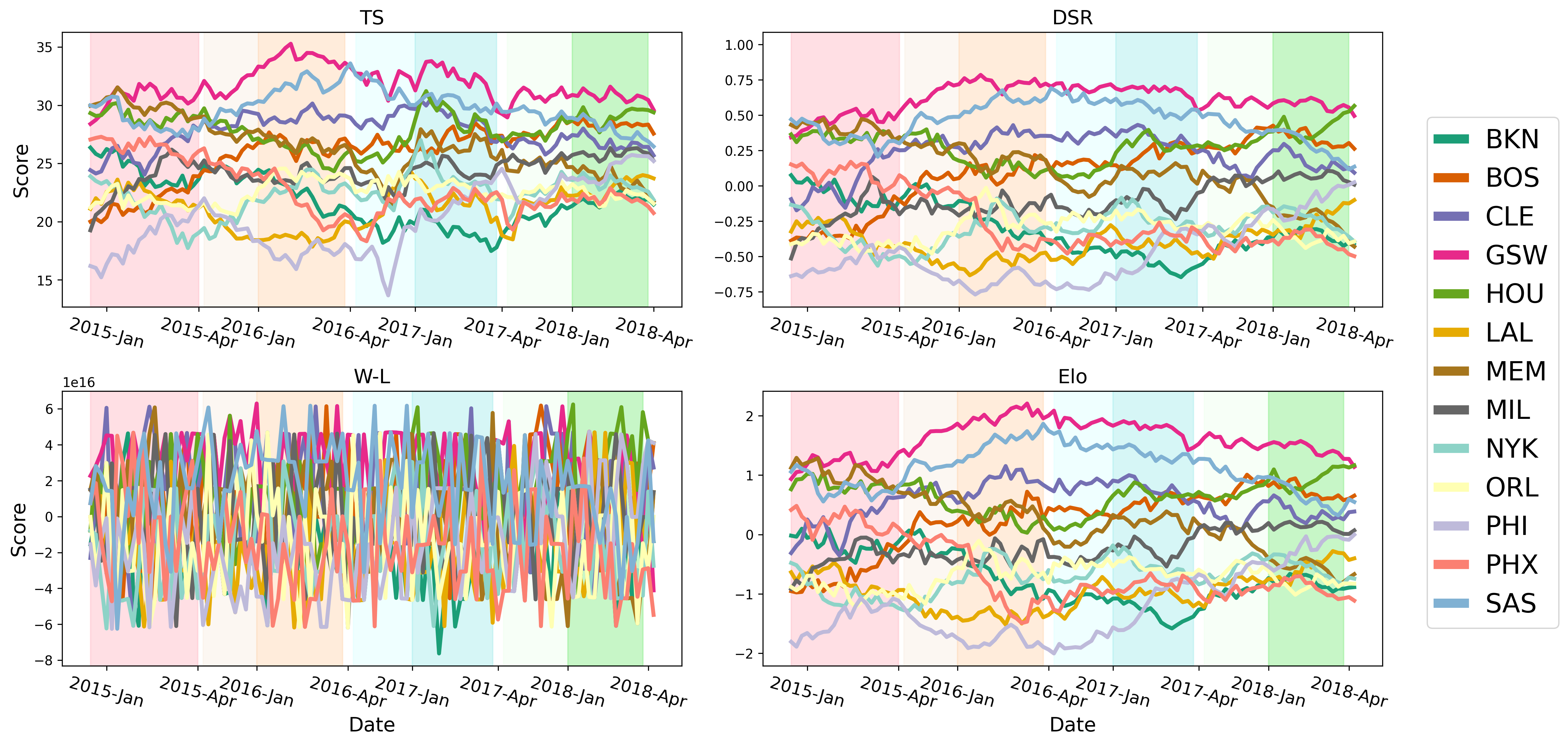}
	\caption{\textbf{Evolution over time of predicted ranks for the NBA dataset.} We illustrate the predicted ranks of four models over time: TS, \dsr, W-L and Elo. We select a subset of 13 teams  (as indicated in the legend) to highlight the behaviors of both top and bottom scoring teams. Vertical colored bands break seasons into two periods.  }
	\label{fig:nba_scores_over_time}
\end{figure*}

We observed qualitative differences of the inferred ranks in \Cref{fig:nba_scores_over_time} similar to those observed in  \Cref{fig:synscores} for synthetic data. W-L infers ranks that change with a much higher frequency than the others. While smoother, the ranks inferred by TS show more frequent variations than \dsr\ and Elo, which infer similarly behaving ranks.  

\subsection{The Relevance of Time}\label{sec:dynamicity}

As a final consideration, we turn to a fundamental question: given a dataset of timestamped interactions, does their chronological order matter? If the answer is positive, then we should use a dynamical ranking algorithm to analyze the data. If not, a simpler static algorithm should be enough.

One way to assess whether a given dataset is better modeled by a dynamical or static algorithm is by randomly permuting the order of interactions---but not their outcomes---and thus removing any relationship between ranks and time. If an algorithm performs significantly better on the original data than on the permuted data, this shows that the order matters and a dynamical model is justified. To be more precise, applying random permutations to the data produces a distribution of any test statistic, including any measure of the performance of an algorithm that predicts which way a given interaction will go (e.g., which of two players will win a chess match, conditioned on the event that they play). If the performance on the original data is far out in the tail of this distribution, we can reject the null hypothesis that the time-steps are simply independent draws from a static model.

We run this permutation test first on synthetic data, confirming as expected that the dynamical model performs significantly better on synthetic data generated with the time-varying model introduced in \Cref{sec:genmod}, provided that the hierarchy itself is sufficiently strong (\Cref{fig:histogram_varying_noise}). However, when the hierarchy is weak (i.e., $\beta$ is small), the ranks have little relationship to the outcomes, and treating the ranks dynamically is no longer justified by the permutation tests (\Cref{fig:histogram_varying_noise}).

\begin{figure*}
	\centering
	\includegraphics[width=\linewidth,height=4.5cm]{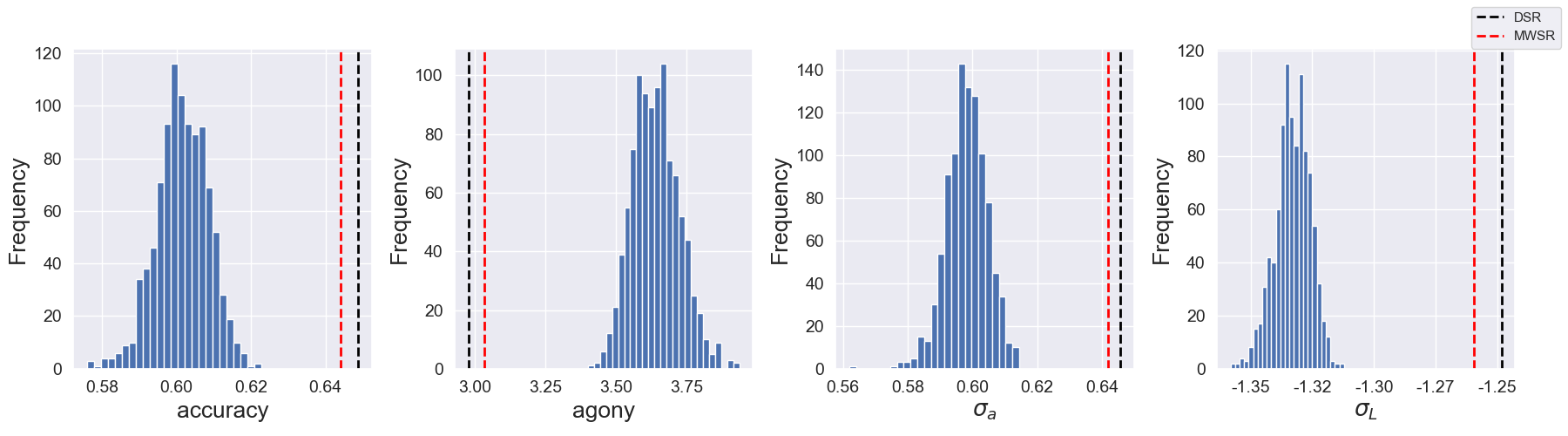}
	\caption{\textbf{Permutation test results on the NBA dataset: chronology matters.} The histogram is generated by $1000$ random permutations to the NBA dataset, and measuring the performance of Dynamical SpringRank on these permuted datasets. The black and red dotted lines represent the results of \dsr and \mwsr\ respectively on the original, chronologically-ordered NBA dataset --- the accuracy is much higher, and the agony much lower, than the vast majority of permuted datasets. This convincingly rejects the null hypothesis that chronological order does not matter, and justifies the use of a dynamical model. In each case the $p$-value is less than $0.001$.}
	\label{fig:histogram_nba}
\end{figure*}

For NBA data, permutation tests show that chronological order matters, and that using a dynamical model significantly improves prediction (\Cref{fig:histogram_nba}). However, for the soccer and chess datasets, we find mixed results depending on the test statistic. For instance, the ``agony'' (a measure which penalizes the model for interactions $i \to j$ if $s_j-s_i$ is large) suggests that time-order is relevant, while the accuracy (the fraction of interactions whose direction is correctly predicted) is less sensitive to this information (\Cref{fig:histogram_real_extra}, \Cref{tb:pvalues}). While the most straightforward explanation is that NBA rankings are more time-varying, while soccer and chess are less so, we also note that there are many more games in a NBA season than in a soccer season, since there are more teams and more frequent games in the NBA, therefore allowing our simple permutation test to reject the null hypothesis more easily with more available data to differentiate time-varying versus static ranks (\Cref{tb:datasets}).

\section*{Conclusion}
\label{sec:conclusions}

\dsrfull\ is a principled extension of the physics-inspired SpringRank model for dynamic hierarchal structures, which lets us infer time-varying ranks from timestamped interactions. By coupling individuals' previous and current ranks, it exploits the chronological ordering of the data to better predict the outcomes of future interactions. It contains a parameter $\kself$ that can be tuned or learned in order to control the smoothness of the change in ranks, or equivalently the weight given to past ranks.

We constructed two different formulations of Dynamic SpringRank: an online and an offline one, which are given just past ranks and the entire history respectively. The online version performed better and is less computationally expensive. However, both models, similar to the static version, are scalable algorithms that require sparse linear algebra and provide a probabilistic generative model for creating dynamically directed networks with tunable levels of hierarchy and sparsity.

We also illustrated that in dynamic settings where time information is important, \dsrfull\ is better than its static counterpart. Its ability to predict future outcomes in dynamical settings proved to be similar or better than other state-of-the-art dynamical ranking algorithms for a variety of metrics and datasets, both synthetic and real. An open-source implementation of both \emph{offline} and \emph{online} versions of \dsrfull\ is available at \href{https://github.com/cdebacco/DynSpringRank}{https://github.com/cdebacco/DynSpringRank}.

For future work, we defined more elaborate models where the time intervals between interactions can vary, or where a momentum term induces smoothness in the rate at which ranks change over time. Another (perhaps challenging) direction is to couple the rank dynamics with the entities' choices to interact with each other. For instance, one can imagine a model in which animals tend to challenge those immediately above them in the dominance hierarchy, or where new arrivals to a community test themselves against current members in order to find their place, or even three-way interactions where an animal who attacks another is punished by a third~\cite{flack2006policing}.  Testing these models would require rich data from biological and social systems.

\section*{Acknowledgements}
We thank Jean-Gabriel Young for his feedback on the boundary conditions of the Offline Dynamical SpringRank model.

\bibliographystyle{apsrev4-1}
\bibliography{bibliography}


\clearpage
\beginsupplement 

\begin{widetext}
	
	\section*{{Appendix (AI)}}
	\label{sec:supp}
	
	\section{Full Derivation Self-Spring Interaction}\label{sec:h_total_derivation}
	
	\noindent Calculate the $i$-th component of the gradient:
	\bea
	\nonumber \frac{\partial \Htotal}{\partial s_i^t}&=&\sum_j [A_{ij}^t(s_i-s_j-\ell_0)-A_{ji}^t(s_j-s_i-\ell_0)] +\kself(s_i^t-s_i^{t-1})\\ 
	\nonumber&=&\sum_j (A_{ij}^t + A_{ji}^t)\,s_i^t-\sum_j (A_{ij}^t + A_{ji}^t)\,s_j^t-\sum_j(A_{ij}^t - A_{ji}^t)\,\ell_0+\kself\,(s_i^t-s_i^{t-1})\\ 
	\nonumber&=&(d_i^{out,t}+d_i^{in,t}+\kself)\,s_i^t-\sum_j (A_{ij}^t + A_{ji}^t)\,s_j^t-(d_i^{out,t}-d_i^{in,t})\ell_0-\kself\,s_i^{t-1} \, .
	\eea
	Imposing $\nabla H=0$ we obtain:
	\be
	(d_i^{out,t}+d_i^{in,t}+\kself)\,s_i^t-\sum_j (A_{ij}^t + A_{ji}^t)\,s_j^t=(d_i^{out,t}-d_i^{in,t})\,\ell_0+\kself\,s_i^{t-1} \, ,\nonumber
	\ee
	which yields:
	\be
	\rup{ D^{out,t}+D^{in,t}- \bup{A^{t}+(A^t)^\dagger}+\kself\id}\,\v{s}^{t,*}=\rup{D^{out,t}-D^{in,t}}\ell_0 \nonumber +k_{0}\, \v{s}^{t-1} \, ,
	\ee
	as reported in~\eqref{eqn:fullsolution} for $\ell_0 = 1$.
	
	\section{Full Derivation of Self-Spring Interaction Over All Time}\label{sec:h_total_derivation_all_time}
	
	\noindent Calculate the $i$-th component of the gradient:
	\begin{align*}
		\frac{\partial \Htotal}{\partial s_i^t}
		&=\frac{\partial}{\partial s_i^t} \left[ \sum_t^T H^t(\boldsymbol{s}^t,\boldsymbol{s}^{t-1}) \right] \\
		&=\frac{\partial}{\partial s_i^t} \left[ \sum_t^T \sum A^t_{ij}H_{ij}(s^t_i,s^t_j) \right] 
		+ \frac{\partial}{\partial s_i^t} \left[ \sum_t^T \sum \kself \Hself(s^t_i,s^{t-1}_i) \right] \\
		&=\sum_{j}^{N}[A^t_{ij}(s^t_i-s^t_j-\ell_0) - A^t_{ji}(s^t_j-s^t_i-\ell_0)]-\kself s^{t-1}_i+2\kself s^t_i-\kself s^{t+1}_i\\
		&=\sum_j (A_{ij}^t + A_{ji}^t)\,s_i^t-\sum_j (A_{ij}^t + A_{ji}^t)\,s_j^t-\sum_j(A_{ij}^t - A_{ji}^t)\ell_0-\kself s_i^{t-1}+2\kself s_i^t -\kself s_i^{t+1}\\
		&=(d_i^{out,t}+d_i^{in,t}+\kself)\,s_i^t-\sum_j (A_{ij}^t + A_{ji}^t)\,s_j^t-(d_i^{out,t}-d_i^{in,t})\ell_0-\kself\,s_i^{t-1} + 2\kself s_i^t -\kself s_i^{t+1}\, .
	\end{align*}
	\\
	Imposing $\nabla H=0$ we obtain:
	\begin{equation*}
		(d_i^{out,t}+d_i^{in,t}+2\kself)\,s_i^t-\sum_j (A_{ij}^t + A_{ji}^t)\,s_j^t=(d_i^{out,t}-d_i^{in,t})\ell_0+\kself(s_i^{t-1}+s_i^{t+1}) \, ,
	\end{equation*}
	which yields:
	\begin{equation*}
		[D^{out,t}+D^{in,t}- (A^{t}+(A^{t})^\dagger)+2\kself \mathbb{I} ]{s}^{t,*}=[D^{out,t}-D^{in,t}]\ell_0  +k_{0} ({s}^{t-1} + {s}^{t+1}) \, ,
	\end{equation*}
	as reported in (\ref{eqn:h_total}) for $\ell_0 = 1$.
	
	\section{Dynamic Spring Rest Length}\label{sec:sidynl}
	As an alternative to the time-dependency presented in the main text (i.e., through self-springs), we also investigated the introduction of a time-dependent rest length. 
	In this case we assume a dynamic rest length $\ell_{ij}^{t}$ for the interaction at time $t$ between $i$ and $j$. To enforce a relationship between current and past ranks, we assume $\ell_{ij}^{t}$ to be a function of the rank difference $s_{i}^{t-1}-s_{j}^{t-1}$ between $i$ and $j$ at time $t-1$:
	\bea
	\nonumber H_{ij}^t\bup{s_i^t,s_j^t}&=&\frac{1}{2}\bup{s_i^t-s_j^t-l_{ij}^{t}}^2 ,
	\eea
	where
	\bea
	l_{ij}^{t}&=&s_i^{t-1}-s_j^{t-1}+\ell_0 \, . \label{eqn:lt}
	\eea
	\\
	The resultant Hamiltonian for the whole system is,
	\bea
	\nonumber H^{t}(\v{s}^t,\v{s}^{t-1})=\sum_{i,j}A_{ij}^{t}H^{t}_{ij}\bup{s_i^t,s_j^t} \, .
	\eea
	As opposed to Eq.~\ref{eqn:fullH}, here we do not have self-interactions. Instead, past ranks appear directly inside the rest lengths.
	If we define a new variable $z_i^t=s_i^t-s_i^{t-1}$,  we obtain the Hamiltonian:
	\bea
	H^{t}(\v{z}^t)=\sum_{i,j}A^{t}_{ij}H^{t}_{ij}\bup{z_i^t,z_j^t}=\sum_{i,j}\frac{A^{t}_{ij}}{2}\bup{z_i^t-z_j^t-\ell_0}^2\nonumber
	\eea
	which is the same Hamiltonian used in static SpringRank \cite{de2018physical} but as a function of the auxiliary variable $\v{z}^{t}$. Thus, we know that the ground state $\v{z}^{t,*}$ will be the solution of the linear system:
	\bea
	\left[ D^{\text{out}}+D^{\text{in}} - \left(A+ A^\dagger \right) \right ] \v{z}_t^* =\left[D^{\text{out}}-D^{\text{in}}  \right ]\ell_0 \ones\, . \nonumber
	\eea
	The idea is that once $\v{z}^{t,*}$ is obtained by solving this linear system, one can extract the ranks as $s_i^t= z_i^t+s_{i}^{t-1}$, where $s_{i}^{t-1}$ is known from the inference of the previous step. Notice that in the extreme case of having only two individuals $i$, $j$, initializing $s_{i}^{0}=s_{j}^{0}=0$ and $i$ as the constant winner ($A_{ij}^{t}\geq 0$ and $A_{ji}^{t}= 0 \, \forall t$), we would infer $s_{i}^{1}-s_{j}^{1}= \ell_0$ at the first time-step. Then iterating in time yields $\ell^t_{ij}= t \ell_0$. In words, for situations where the hierarchy is strong and time is constant (i.e., a stronger individual always defeats a weaker one at any time-step), the rest length would grow linearly in time. As a consequence, the distance between ranks grows further and further, driving them apart. This is the case in sports, for instance, where teams earn points for each win, distancing them more and more from the losing teams. In other situations, we might want instead a scenario where the difference between ranks becomes a constant value $\ell_0$ the more we collect consistent observations in time, i.e., $\forall t, s_{i}^{t}-s_{j}^{t}=\ell_0$. This can be easily obtain by changing the model's details, like setting a different initial rest length and update in Eq.~(\ref{eqn:lt}).
	
	\section{Performance Evaluation}\label{sisec:evaluation}
	In this section, we discuss the various metrics used in more detail.
	Accuracy is a coarse-grained measure to evaluate the quality of predictions. It is the fraction of times an observed directed edge points from the higher towards the lower ranked node, i.e., the number of times that a \emph{stronger} (according to our ranking) individual \emph{beats} a weaker one,
	\be
	\nonumber 
	\textrm{accuracy} = \f{1}{M} \sum_{i,j} A_{ij} \, \Theta(s_i - s_j) \, ,
	\ee
	where $\Theta(x)=1$ if $x > 0$, $\Theta(x)=0.5$ if $x = 0$ and $\Theta(x)=0$ if $x < 0$; $M=\sum_{i,j} A_{ij}$. 
	
	If we call an \emph{upset} an interaction where a lower ranked individual beats someone stronger, then the accuracy is just 1 minus the fraction of upsets. Accuracy does not weigh upsets differently. However, in certain situations making an erroneous prediction involving individuals nearby in rank might be less important than an error involving individuals far in rank. In this case, it is useful to consider the \emph{agony} function \cite{gupte2011finding}. It considers the difference in ordinal ranks as penalties\footnote{We use \emph{positional} ranks instead of the real-valued ranks to avoid scale problems comparing different algorithms}. Subsequently, an upset between two nodes close in rank counts much less than an upset between two nodes far rank, based on a parameter $d$:
	\begin{equation*}
		\textrm{agony} = \frac{1}{M}\sum_{i,j} A_{ij} \hspace{0.1cm}  max(0,r_i - r_j)^d \, ,
	\end{equation*}
	where $r_i\in [0,..,n-1]$ is the \emph{ordinal} rank of node $i$ (which can naturally be extracted from the real-valued ranks $s_{i}$). When $d = 0$ we recover the standard number of unweighted upsets. We use $d=1$ in our evaluation of models. The more the rank is informative towards the predicted outcomes, the lower the value of the agony and the less the hierarchy is violated.

	Accuracy and agony are metrics for ordinal rankings. For real-valued models such as SpringRank, it is worth considering fine-grained metrics as well.
	We thus consider in our experiments two other metrics that take into account an estimate of $P_{ij}$ -- the probability that $i$ beats $j$.
	
	First, $\sigma_a$ is the average probability assigned to the correct direction of an edge:
	\begin{equation*}
		\sigma_a= 1 - \frac{1}{2M}\sum_{ij}\vert A_{ij}-\overline{A_{ij}}P_{ij}\vert \, ,
	\end{equation*}
	where $\overline{A_{ij}}=A_{ij}+A_{ji}$ is the number of interactions between $i$ and $j$.
	
	Second, $\sigma_L$ is the conditional log-likelihood of generating the directed edges \emph{given} their existence:
	\begin{align*}
		\sigma_L&=\log P\bup{A|\bar{A}}\\
		&=\sum_{ij}  \binom{A_{ij}+A_{ji}}{A_{ij}}+\log\left[P_{ij}(\beta)^{A_{ij}}\bup{1-P_{ij}(\beta)}^{A_{ji}}\right]. \nonumber
	\end{align*}
	
	Notice that we explicitly highlight the dependence of $P_{ij}$ on the (inverse) \emph{temperature} parameter $\beta$ which control the level of hierarchy in the predictions. For $\beta \rightarrow \infty$ the network is fully hierarchical which means that an edge between $i$ and $j$, with $s_{i}>s_{j}$, points from $i\rightarrow j$ with $P_{ij}=1$. In contrast, when $\beta=0$, the predicted outcomes are completely random with $P_{ij}=P_{ji}=0.5$. 
	
	In general, maximizing $\sigma_a$ and $\sigma_L$ requires two distinct values for $\beta$ that we will denote as $\hat{\beta}_a$ and $\hat{\beta}_L$. Intuitively, the reason is that a single severe mistake where $A_{ij}=1$ but $P_{ij} \approx 0$ reduces the likelihood by a large amount, while only reducing the accuracy by one edge.  As a result, predictions using $\hat{\beta}_a$ produce fewer incorrectly oriented edges and achieve a higher $\sigma_a$ on the test set. On the other hand, predictions using $\hat{\beta}_L$ will produce fewer dramatically incorrect predictions where $P_{ij}$ is very low, and thus achieve higher $\sigma_L$ on the test set \cite{de2018physical}. In other words, a prediction model that maximizes $\sigma_L$ tends to be more cautious in assigning high probabilities of \emph{success}, even in very unbalanced matches, in order to avoid potential impactful mistakes. In contrast, a model optimizing $\sigma_a$ can be less conservative, ignoring isolated (even dramatic) mistakes and favoring a good frequency of predictions as close as possible to the real probability.

	\section{Description of algorithms used for comparison}

	\label{apx:description_algs}
	\noindent\textbf{Elo rating system \cite{elo1978rating}.} This method assumes an hidden score $R_{i}$ for each node $i$. The expected score $S_{ij}$ of a game between players $i$ and $j$ is a function of the score difference $R_{i}-R_{j}$ as: 
	\be
	S_{ij}(R_{i},R_{j})= \frac{1}{1+10^{-(R_i-R_j)/400}}\quad.
	\ee

	The actual score $A_{ij}$ of the game is $1$ if player $i$ wins, $1/2$ if the game is a draw, and $0$ if player $i$ loses. After observing the outcome of the match, the score of $i$ is updated according to the following rule:
	\be
	R_i^{new}=R_i + K(A_{ij}-S_{ij}) \quad,
	\ee
	where $K$ is  an attenuation factor that determines the weight that should be given to a player’s performance relative to their previous rating. We used grid-search to determine $K$. The above formula has a natural interpretation. The term $A_{ij}-S_{ij}$ represents a discrepancy between what was expected and what was observed. If this term is positive, then the player achieved a result better than what predicted by the rating at the previous time step. Hence, the player’s rating is increased to reflect the possible improvement in strength. Similarly, if the term  $A_{ij}-S_{ij}$ is negative, then the player performed worse than expected. Hence,  this player’s rating decreases by the discrepancy magnified by the value $K$. 
		\\~\\
		
	\textbf{WHR system \cite{coulom2008whole}.} This algorithm is based on the dynamic Bradley-Terry model \cite{glickman1993paired}. The Bradley-Terry model for paired comparisons assumes that each node $i$ has a rating $\gamma_i(t)=10^{R_i(t)/400}$, where $R_i(t)$ is the Elo rating of player $i$ at time $t$. Based on this, the probability of $i$ winning a game against $j$ at time $t$ is:
	\be\label{eqn:BTL}
	P(A_{ij}^{t}>0|\gamma_{i},\gamma_{j})=\frac{\gamma_i(t)}{\gamma_i(t)+\gamma_j(t)}\quad.
	\ee
	The WHR algorithm consists in estimating the values of $\gamma(t)$ using posterior inference of $p(\gamma|A)$ via Bayes' rule using the following expression:
	\be
	p(\gamma|A)=\frac{P(A|\gamma)\, p(\gamma)}{P(A)}\quad,
	\ee
	where $p(\gamma)$ is a prior distribution on $\gamma$, $P(A)$ is a normalizing constant and $P(A|\gamma)$ is the Bradley-Terry model described in \cref{eqn:BTL}. 
	\\~\\
	\textbf{Dynamic win-lose score \cite{motegi2012network}.} This method assumes two scores for each node $i$ at any given time step $t_{n}$, a win score $w_{t_{n},i}$ and a loss score $\ell_{t_{n},i}$. Let $A_{t}$ be the win-loss matrix for the game that occurs at time $t_n$ $(1\leq n \leq n_{max})$. If player $j$ wins against player $i$ at time $t_n$, the $(i,j)$ element of the matrix $A_{t_n}$ is set to $1$. All the other elements of $A_{t_n}$ are set to $0$. The method accounts for the effect of wins or losses by using a discounted past history and indirect results, i.e. results involving players that compete against a common opponent. Formally, it defines a ``win'' matrix $W_{t_n}$ as follows:
	\begin{align*}
		W_{t_n}= & A_{t_n}+e^{-\beta\left(t_n-t_{n-1}\right)} \sum_{m_n \in\{0,1\}} \alpha^{m_n} A_{t_{n-1}} A_{t_n}^{m_n} \\
		& +e^{-\beta\left(t_n-t_{n-2}\right)} \sum_{m_{n-1}, m_n \in\{0,1\}} \alpha^{m_{n-1}+m_n} A_{t_{n-2}} A_{t_{n-1}}^{m_{n-1}} A_{t_n}^{m_n} \\
		& +\cdots+e^{-\beta\left(t_n-t_1\right)} \sum_{m_2, \ldots, m_n \in\{0,1\}} \alpha^{\sum_{i=2}^n m_i} A_{t_1} A_{t_2}^{m_2} \cdots A_{t_n}^{m_n} \quad,
	\end{align*}
	where $\alpha$ is the weight of an indirect win and $\beta\geq 0$ represents the decay rate of the score in time. These are the two main hyperparameters of this model, we fix them using cross-validation. The first term $A_{t_n}$ on the right-hand side represents the effect of the direct win at time $t_n$. The second term consists of two contributions. For $m_n=0$, the quantity inside the sum represents the direct win at time $t_{n-1}$, which results in weight $e^{-\beta(t_n-t_{n-1})}$, discounted depending on $\beta$ and the time passed between two time steps. For $m_n=1$, the quantity represents the indirect win. The $(i, j)$ element of $A_{t_{n-1}} A_{t_n}$ is positive if and only if player $j$ wins against a player $k$ at time $t_n$ and $k$ wins against $i$ at time $t_{n-1}$. Player $i$ gains score $e^{-\beta(t_n-t_{n-1})}$ $\alpha$ out of this situation. For both cases $m_n=0$ and $m_n=1$,the $j$-th column of the second term accounts for the effect of the $j$'s win at time $t_{n-1}$. The other terms behave analogously considering also third order indirect interactions and so on. A similar matrix is defined to account for losses. Then, the win score $w_{t_{n},i}$ of a player $i$ is computed as the $i$-th entry of the vector $w_{t_{n}}=W^{T}_{t_{n}}\mathbf{1}$, where $\mathbf{1}$ is the all-one vector. Similarly, one can get the loss score $\ell_{t_{n},i}$ by considering the loss matrix. The final score for a player $i$ at time $t_{n}$ is the difference $s_{t_{n},i}=w_{t_{n},i}-\ell_{t_{n},i}$.\\
	\newline
	\textbf{TrueSkill rating system \cite{dangauthier2008trueskill}.}  TrueSkill’s current belief about a player’s skill $s_{i,t}$ at time $t$ is represented by a Gaussian distribution with mean $\mu_{i}$ and variance $\sigma_{i}^2$. This is inferred using Bayesian inference, where the goal is to estimate the posterior distribution $P(s_{i,t}|r)$, where $r$ is a vector containing the rank of the nodes as determined by the outcomes, i.e. is a quantity determined by input data. The influence of the skill at the previous time step enters as a Gaussian prior centred at the value of the skill at the previous time step, i.e. $P(s_{i,t}|s_{i,t-1},\gamma^{2})= \mathcal{N}(s_{i,t};s_{i,t-1},\gamma^{2})$, where $\gamma$ is an hyperparameter. This method estimates the posterior distributions $P(s_{i,t}|r)$ with a Bayesian inference procedure that performs a Gaussian filtering that repeatedly smoothes the scores forward and backward in time, using approximate message-passing.

	\section{Cross-Validation and Hyperparameter Tuning}\label{sisec:tuning}
	We provide more technical details about the hyperparameter tuning used in the various algorithms and experiments. In all cases, we assume training and test sets have a chronological order, i.e., all matches in the train set happen earlier than those in the test set. Regardless of hyperparameters, all cross-validation folds provide the same exact train/test set to each algorithm for a fair comparison. Importantly, test sets are only used for evaluation.
	
	All results displayed were computed with cross-validation which entailed using 50\% of the total data as a train set and 4 time-steps as a test set. This interval was shifted by 1 time-step each fold. Fig~\ref{fig:cross_val} demonstrates this process. As a result of cross-validation, there are at most four different values for the same time-step. The reported results are an average of these values.
	\begin{figure}[hbt!]
		\includegraphics[width=11cm,height=4cm]{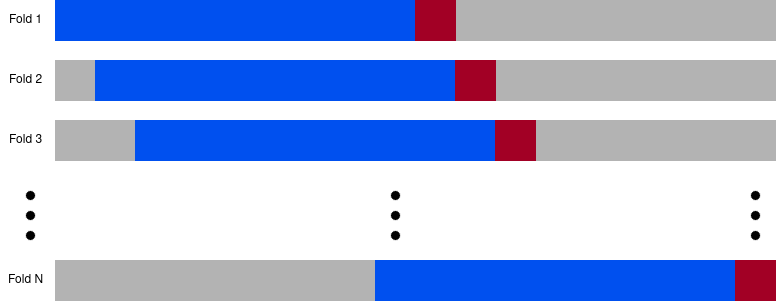}
		\caption{\textbf{Illustration of the cross-validation used in experiments.} The blue bar represents the training set, the red bar is the test and the grey is the total dataset. N is the total number of folds.}
		\label{fig:cross_val}
	\end{figure}
	
	As previously mentioned, we used grid-search to perform hyperparameter tuning. For \dsrfull, grid-search is divided into two steps: first, finding the order of magnitude of $\kself$ and then progressively finding a more precise value. Refer to Algorithm \ref{alg:grid_search} for the pseudocode of the procedure followed.
	\begin{algorithm}[hbt!]
		\caption{Grid-Search}
		\label{alg:grid_search}
		\begin{algorithmic}[1]
			\STATE $K =\{0.001, 0.01, 0.1, 1, 10, 100, 1000\}$
			\FOR{$i \gets -1,-2,-3$}
			\FOR{$\kself$ in $K$}\\
			\quad \quad Find $\kself^{*}$, the optimal $\kself$ that produces best result
			\ENDFOR
			\STATE Update interval $K = [\kself^{*} - 10^i,\quad \kself^{*} + 10^i]$
			\ENDFOR
		\end{algorithmic}
	\end{algorithm}
	
	Three algorithms (\mwsr, TS and WHR) require an optimal window size, $\tau_{\opt}$, for storing data in the training set. We chose this by varying the window size, calculating the average value for each performance metric inside the training set and then choosing the window size corresponding to the best of each of these values. Since the reported results are due to cross-validation, on average the window size of  \mwsr, TS and WHR on the NBA dataset is $\tau_{\opt}=13, 23, 31$ respectively.
	
	Next, Elo requires a scaling factor $k$ which was determined through a grid-search in the interval $[0.01,1)$. WL requires a decaying factor $\beta=3$ and a weighting for indirect wins $\alpha=0.005$, both of which were fixed with cross-validation. Finally, there are different versions of static SpringRank and we considered the standard version with regularization $\alpha=0$.
	
	\section{Implementation of \nmdsr and its Boundary Conditions}\label{apx:offdsr_bc}
	In our experiments we created a realistic scenario in which each algorithm had to predict the future ranks of a node given its past interactions, as if the predictions were taking place in ``real-time''. Thus, only past information was given to each model. \nmdsrfull\ depends on both past and future ranks (see \cref{eqn:h_total}). As a consequence of our experimental choice, the information it was given during the experiments was restricted to only the past. However, the description of \nmdsr\ in \Cref{sec:model} relates more to a scenario where all information is available and ranks are inferred in hindsight.
	
	Next, here we further discuss the boundary conditions of \nmdsr. Our choice for the boundary conditions on the ranks is implemented by removing the following terms from \cref{eqn:h_total}: $\v{s}^{t-1}$ and $\v{s}^{t+1} $for $t=0$ and $t=T$ respectively. This is equivalent to $\v{s}^{-1} = 0 =\v{s}^{T+1} $. The effect of this boundary condition is that ranks close to the boundary conditions are slightly pulled towards zero. This has a greater influence on datasets with a small number of timesteps. However, the effect lessens with more timesteps.  Alternate boundary conditions may be chosen, such as $\v{s}^{-1} =\v{s}^{0}$  and $ \v{s}^{T}  =\v{s}^{T+1} $.  The effect of two aforementioned boundary conditions on a toy example with only two nodes and one directed edge between them is illustrated in \Cref{fig:offdsr_bc_zero} and \Cref{fig:offdsr_bc_previous_step}, respectively. We do not explore the effects of different boundary conditions in our experiments and leave it for future work.
	
	\begin{figure}[!h]
		\centering
		{\includegraphics[width=0.7\linewidth,height=14cm]{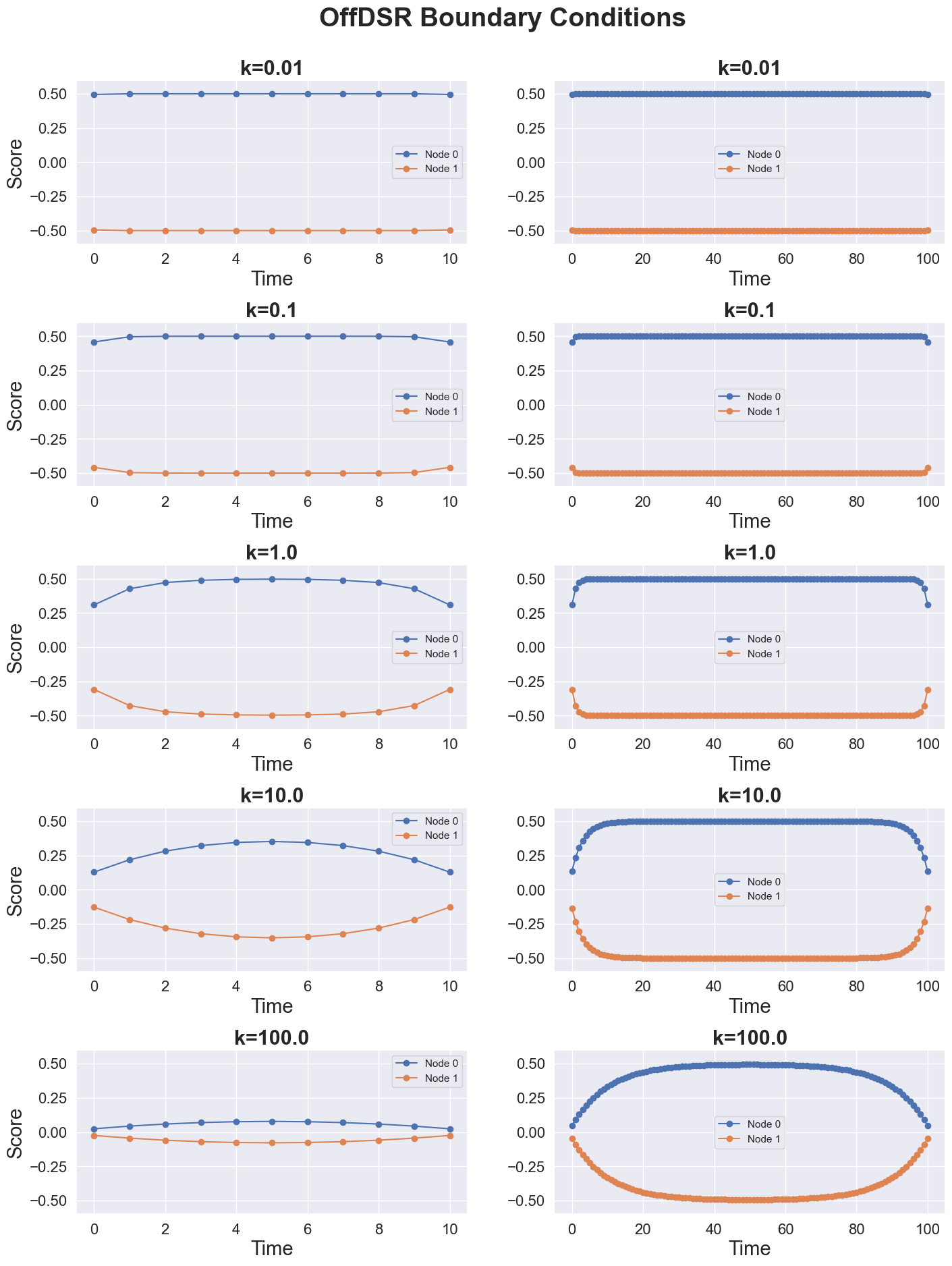}}
		\caption{\textbf{\nmdsr with boundary conditions set to zero.} A toy example was used to generate the plot where two nodes interact with a single directed edge pointing in the same direction for every timestep. The parameter $k$ of \nmdsr is varied as well as the number of timesteps in order to further illustrate the effect of the boundary conditions on the ranks.}
		\label{fig:offdsr_bc_zero}
	\end{figure}
	\begin{figure}[!h]
		{\includegraphics[width=0.7\linewidth,height=14cm]{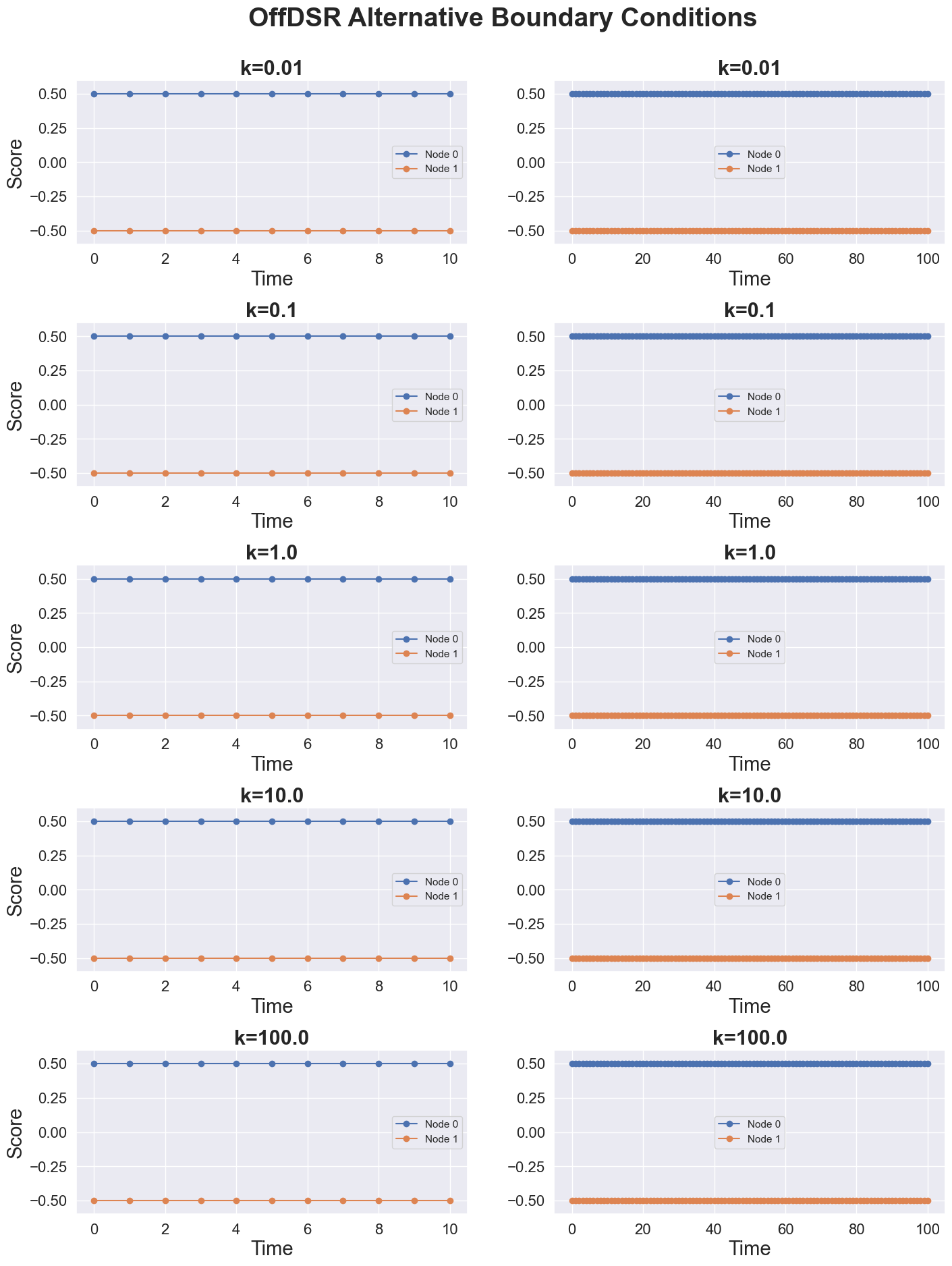}}
		\caption{\textbf{ \nmdsr with alternative boundary conditions.} The first boundary condition is $\v{s}^{t-1}= \v{s}^{t}$ where $t=0$ and the second boundary condition is $\v{s}^{t}=\v{s}^{t+1} $where $t=T$. A toy example was used to generate the plot where two nodes interact with a single directed edge pointing in the same direction for every timestep. This is the same as in \cref{fig:offdsr_bc_zero}. The parameter $k$ of \nmdsr is varied as well as the number of timesteps in order to further illustrate the effect of the boundary conditions on the ranks.}
		\label{fig:offdsr_bc_previous_step}
	\end{figure}
	
	\section{Synthetic Experiments}\label{apx:synthetic_experiments}
	
	\paragraph*{Periodic Evolution of Synthetic Ranks.}We consider a periodic evolution of the ranks generated for synthetic experiments, expressed as Eq.~\eqref{eq:dyn_score}. To add detail to the extraction process of the parameters, they were selected from a continuous uniform distribution. The interval of the distribution for parameters was as follows: $b_{i},c_{i}\in[-1,1)$, as it is the standard range for a cosine function;  $\omega_i,\upsilon_i\in[-1,2)$  in order to vary the frequency with which scores change, with larger increases reflected in values between 1 and 2 being less likely than values between -1 and 1; finally, $\phi_i\in[0,1)$ to ensure that scores do not have the same rate of change at the beginning of the time interval.
	\paragraph*{Standard errors.} We report standard errors on synthetic experiments where we vary the noise level represented by the parameter $\beta$ in \Cref{tb:sem_beta}. These complement \Cref{tab:varying_noise}  in the main manuscript.
	
	\begin{table}[h!]
		\centering
		\begin{tabular}{cc|cccccccc}
			\textbf{$\beta$} & \textbf{Metric}         &     \textbf{Elo} &  \textbf{\nmdsr} &    \textbf{\mwsr} &     \textbf{\dsr} &      \textbf{SR} &      \textbf{TS} &       \textbf{W-L} &     \textbf{WHR} \\
			\hline
			\multirow{4}*{\textbf{0.1}}  & accuracy &  0.0073 &  0.0065 &  0.0067 &  0.0070 &  0.0064 &  0.0068 &  0.0068 &  0.0077 \\
			& agony &  0.0286 &  0.0269 &  0.0278 &  0.0267 &  0.0331 &  0.0265 &   0.0325 &  0.0284 \\
			& $\sigma_a$ &  0.0028 &  0.0031 &  0.0044 &  0.0039 &  0.0032 &  0.0029 &        -- &  0.0028 \\
			& $\sigma_L$ &  0.0107 &  0.0008 &  0.0055 &  0.0044 &  0.0019 &  0.0077 &        -- &  0.0066 \\
			\hline
			\multirow{4}*{\textbf{0.5}}  & accuracy &  0.0056 &  0.0054 &  0.0057 &  0.0056 &  0.0042 &  0.0051 &  0.0074 &  0.0054 \\
			& agony &  0.0197 &  0.0196 &  0.0202 &  0.0203 &  0.0168 &  0.0190 &  0.0314 &  0.0185 \\
			& $\sigma_a$ &  0.0027 &  0.0030 &  0.0051 &  0.0048 &  0.0031 &  0.0029 &        -- &  0.0029 \\
			& $\sigma_L$ &  0.0215 &  0.0035 &  0.0127 &  0.0114 &  0.0061 &  0.0121 &        -- &  0.0110 \\
			\hline
			\multirow{4}*{\textbf{1.0}} & accuracy &  0.0056 &  0.0053 &  0.0048 &  0.0050 &  0.0073 &  0.0047 &  0.0060 &  0.0044 \\
			& agony &  0.0164 &  0.0156 &  0.0152 &  0.0161 &  0.0297 &  0.0141 &  0.0218 &  0.0142 \\
			& $\sigma_a$ &  0.0035 &  0.0037 &  0.0044 &  0.0047 &  0.0059 &  0.0030 &        -- &  0.0029 \\
			& $\sigma_L$ &  0.0318 &  0.0079 &  0.0153 &  0.0159 &  0.0254 &  0.0151 &        -- &  0.0135 \\
			\hline
			\multirow{4}*{\textbf{1.5}} & accuracy &  0.0049 &  0.0050 &  0.0050 &  0.0048 &  0.0052 &  0.0049 &  0.0062 &  0.0046 \\
			& agony &  0.0126 &  0.0132 &  0.0124 &  0.0131 &  0.0210 &  0.0129 &  0.0243 &  0.0118 \\
			& $\sigma_a$ &  0.0035 &  0.0040 &  0.0046 &  0.0043 &  0.0047 &  0.0037 &        -- &  0.0034 \\
			& $\sigma_L$ &  0.0346 &  0.0080 &  0.0203 &  0.0232 &  0.0263 &  0.0187 &        -- &  0.0148 \\
			\hline
			\multirow{4}*{\textbf{2.0}}  & accuracy &  0.0038 &  0.0042 &  0.0036 &  0.0040 &  0.0053 &  0.0047 &  0.0056 &  0.0046 \\
			& agony &  0.0092 &  0.0095 &  0.0080 &  0.0082 &  0.0206 &  0.0094 &  0.0183 &  0.0089 \\
			& $\sigma_a$ &  0.0032 &  0.0034 &  0.0035 &  0.0038 &  0.0051 &  0.0034 &        -- &  0.0031 \\
			& $\sigma_L$ &  0.0301 &  0.0073 &  0.0152 &  0.0144 &  0.0302 &  0.0163 &        -- &  0.0147 \\
		\end{tabular}
		\caption{\textbf{Standard error of results from synthetic data with varying noise levels, represent by $\beta$}}
		\label{tb:sem_beta}
	\end{table}

	\paragraph*{Results for Varying Network Density.}
	In \Cref{tab:varying_density,tb:sem_density} , we show results on synthetic data where we vary the network density represented by the parameter $c$.
	
	\begin{table}[!htb]
		\resizebox{11cm}{3.8cm}{
			\begin{tabular}{c|c|cccccccc}
				{$\boldsymbol{c}$} & {\textbf{Metric}} & {\textbf{Elo}} & {\textbf{\nmdsr}} & {\textbf{\mwsr}} & {\textbf{\dsr}} & {\textbf{SR}} & {\textbf{TS}} & {\textbf{W-L}} & {\textbf{WHR}} \\
				\hline
				\multirow[t]{4}{*}{\textbf{1.0}} & accuracy & 0.905 & 0.901 & 0.903 & 0.904 & 0.774 & \sethlcolor{green}\hl{\textbf{0.905}} & 0.845 & 0.903 \\
				\rowcolor[gray]{0.95}[\tabcolsep][\tabcolsep]
				& agony & 0.161 & 0.167 & 0.163 & 0.165 & 0.562 & \sethlcolor{green}\hl{\textbf{0.159}} & 0.299 & 0.162 \\
				& $\sigma_a$ & 0.895 & 0.892 & 0.909 & \sethlcolor{green}\hl{\textbf{0.910}} & 0.778 & 0.889 & -- & 0.884 \\
				\rowcolor[gray]{0.95}[\tabcolsep][\tabcolsep]
				& $\sigma_L$ & -0.574 & -0.705 & -0.459 & -0.459 & -1.054 & -0.451 & -- & \sethlcolor{green}\hl{\textbf{-0.450}} \\
				\hline
				\multirow[t]{4}{*}{\textbf{1.5}} & accuracy & 0.900 & 0.897 & 0.902 & \sethlcolor{green}\hl{\textbf{0.903}} & 0.770 & 0.902 & 0.852 & 0.903 \\
				\rowcolor[gray]{0.95}[\tabcolsep][\tabcolsep]
				& agony & 0.161 & 0.171 & 0.159 & \sethlcolor{green}\hl{\textbf{0.157}} & 0.608 & 0.162 & 0.280 & 0.157 \\
				& $\sigma_a$ & 0.899 & 0.902 & 0.908 & \sethlcolor{green}\hl{\textbf{0.909}} & 0.771 & 0.894 & -- & 0.895 \\
				\rowcolor[gray]{0.95}[\tabcolsep][\tabcolsep]
				& $\sigma_L$ & -0.581 & -0.617 & -0.462 & -0.459 & -1.155 & -0.460 & -- & \sethlcolor{green}\hl{\textbf{-0.452}} \\
				\hline
				\multirow[t]{4}{*}{\textbf{2.0}} & accuracy & \sethlcolor{green}\hl{\textbf{0.909}} & 0.905 & 0.904 & 0.907 & 0.768 & 0.904 & 0.874 & 0.904 \\
				\rowcolor[gray]{0.95}[\tabcolsep][\tabcolsep]
				& agony & \sethlcolor{green}\hl{\textbf{0.147}} & 0.156 & 0.154 & 0.148 & 0.601 & 0.154 & 0.217 & 0.153 \\
				& $\sigma_a$ & 0.911 & 0.910 & 0.915 & \sethlcolor{green}\hl{\textbf{0.916}} & 0.772 & 0.907 & -- & 0.902 \\
				\rowcolor[gray]{0.95}[\tabcolsep][\tabcolsep]
				& $\sigma_L$ & -0.579 & -0.575 & -0.464 & -0.453 & -1.152 & \sethlcolor{green}\hl{\textbf{-0.452}} & -- & -0.453 \\
				\hline
				\multirow[t]{4}{*}{\textbf{2.5}} & accuracy & 0.904 & 0.904 & 0.905 & \sethlcolor{green}\hl{\textbf{0.906}} & 0.763 & 0.905 & 0.877 & 0.905 \\
				\rowcolor[gray]{0.95}[\tabcolsep][\tabcolsep]
				& agony & 0.159 & 0.160 & 0.159 & \sethlcolor{green}\hl{\textbf{0.155}} & 0.601 & 0.160 & 0.216 & 0.160 \\
				& $\sigma_a$ & 0.912 & 0.913 & 0.916 & \sethlcolor{green}\hl{\textbf{0.918}} & 0.773 & 0.909 & -- & 0.907 \\
				\rowcolor[gray]{0.95}[\tabcolsep][\tabcolsep]
				& $\sigma_L$ & -0.601 & -0.650 & -0.470 & -0.463 & -1.172 & -0.459 & -- & \sethlcolor{green}\hl{\textbf{-0.458}} \\
				\hline
				\multirow[t]{4}{*}{\textbf{3.0}} & accuracy & \sethlcolor{green}\hl{\textbf{0.910}} & 0.908 & 0.908 & 0.909 & 0.768 & 0.909 & 0.886 & 0.909 \\
				\rowcolor[gray]{0.95}[\tabcolsep][\tabcolsep]
				& agony & \sethlcolor{green}\hl{\textbf{0.147}} & 0.150 & 0.152 & 0.149 & 0.605 & 0.152 & 0.198 & 0.151 \\
				& $\sigma_a$ & 0.921 & 0.921 & 0.920 & \sethlcolor{green}\hl{\textbf{0.921}} & 0.776 & 0.917 & -- & 0.915 \\
				\rowcolor[gray]{0.95}[\tabcolsep][\tabcolsep]
				& $\sigma_L$ & -0.549 & -0.559 & -0.455 & -0.447 & -1.158 & -0.438 & -- & \sethlcolor{green}\hl{\textbf{-0.438}} \\
		\end{tabular}}
		\caption{\textbf{Results obtained from synthetic data with varying density levels, represented by $\boldsymbol{c}$.} Each value is the mean of 4 independent realizations of the model. The green highlighted values are the top performances for the considered metric. Notably, some of the values in the same row appear identical but only a single value is highlighted. This is because the highlighted value is better by less than three decimal places. \Cref{tb:sem_density} contains the standard error for the above values. $\sigma_a$ and $\sigma_L$ cannot be applied to the W-L model hence there are no values for the metrics.}
		\label{tab:varying_density}
	\end{table}
	
	\begin{table}[h!]
		\centering
		\begin{tabular}{cc|cccccccc}
			\textbf{c} & \textbf{Metric}         &     \textbf{Elo} &  \textbf{\nmdsr} &    \textbf{\mwsr} &     \textbf{\dsr} &      \textbf{SR} &      \textbf{TS} &       \textbf{W-L} &     \textbf{WHR} \\
			\hline
			\multirow{4}*{\textbf{1.0}} & accuracy &  0.0021 &  0.0025 &  0.0023 &  0.0024 &  0.0040 &  0.0024 &  0.0040 &  0.0022 \\
			& agony &  0.0044 &  0.0055 &  0.0048 &  0.0050 &  0.0138 &  0.0052 &  0.0093 &  0.0051 \\
			& $\sigma_a$ &  0.0017 &  0.0023 &  0.0020 &  0.0020 &  0.0031 &  0.0021 &        -- &  0.0021 \\
			& $\sigma_L$ &  0.0171 &  0.0087 &  0.0099 &  0.0099 &  0.0200 &  0.0106 &        -- &  0.0095 \\
			\hline
			\multirow{4}*{\textbf{1.5}} & accuracy &  0.0026 &  0.0027 &  0.0025 &  0.0023 &  0.0039 &  0.0026 &  0.0034 &  0.0026 \\
			& agony &  0.0051 &  0.0056 &  0.0053 &  0.0050 &  0.0180 &  0.0057 &   0.0086 &  0.0056 \\
			& $\sigma_a$ &  0.0016 &  0.0019 &  0.0021 &  0.0020 &  0.0034 &  0.0016 &        -- &  0.0017 \\
			& $\sigma_L$ &  0.0151 &  0.0053 &  0.0084 &  0.0085 &  0.0296 &  0.0086 &        -- &  0.0085 \\
			\hline
			\multirow{4}*{\textbf{2.0}}  & accuracy &  0.0019 &  0.0021 &  0.0020 &  0.0019 &  0.0032 &  0.0022 &   0.0025 &  0.0023 \\
			& agony &  0.0047 &  0.0044 &  0.0049 &  0.0046 &  0.0130 &  0.0050 &   0.0052 &  0.0052 \\
			& $\sigma_a$ &  0.0015 &  0.0018 &  0.0019 &  0.0018 &  0.0026 &  0.0016 &        -- &  0.0016 \\
			& $\sigma_L$ &  0.0176 &  0.0079 &  0.0095 &  0.0087 &  0.0273 &  0.0095 &        -- &  0.0089 \\
			\hline
			\multirow{4}*{\textbf{2.5}} & accuracy &  0.0017 &  0.0016 &  0.0016 &  0.0017 &  0.0036 &  0.0017 &  0.0021 &  0.0018 \\
			& agony &  0.0042 &  0.0040 &  0.0040 &  0.0039 &  0.0140 &  0.0045 &  0.0044 &  0.0046 \\
			& $\sigma_a$ &  0.0014 &  0.0013 &  0.0015 &  0.0014 &  0.0030 &  0.0014 &        -- &  0.0014 \\
			& $\sigma_L$ &  0.0143 &  0.0065 &  0.0076 &  0.0073 &  0.0285 &  0.0076 &        -- &  0.0072 \\
			\hline
			\multirow{4}*{\textbf{3.0}}  & accuracy &  0.0015 &  0.0018 &  0.0016 &  0.0017 &  0.0027 &  0.0017 &  0.0022 &  0.0016 \\
			& agony &  0.0031 &  0.0037 &  0.0036 &  0.0035 &  0.0146 &  0.0035 &  0.0051 &  0.0037 \\
			& $\sigma_a$ &  0.0010 &  0.0013 &  0.0014 &  0.0013 &  0.0025 &  0.0010 &        -- &  0.0010 \\
			& $\sigma_L$ &  0.0102 &  0.0046 &  0.0061 &  0.0059 &  0.0278 &  0.0055 &        -- &  0.0054 \\
		\end{tabular}
		\caption{\textbf{Standard error of results from synthetic data with varying density, represented by c.}}
		\label{tb:sem_density}
	\end{table}

	\paragraph*{Synthetic Ranks in static scenarios.} We consider  static ranks $s_{i}^{t}=s_{i}$ generated synthetically using \Cref{eq:dyn_score} as a sanity check of our permutation test for model selection between static and dynamic models. Results are shown in \Cref{tab:static_results} and \Cref{tb:sem_static}.
	
	\begin{table}[h]
		\resizebox{0.5\linewidth}{!}{
			\begin{tabular}{c|cccccccc}
				{\textbf{Metric}} & {\textbf{Elo}} & {\textbf{\nmdsr}} & {\textbf{\mwsr}} & {\textbf{\dsr}} & {\textbf{SR}} & {\textbf{TS}} & \textbf{{W-L}} & {\textbf{WHR}} \\
				\hline
				\textbf{accuracy} & 0.715 & 0.716 & 0.696 & 0.720 & 0.722 & \sethlcolor{green}\hl{\textbf{0.724}} & 0.552 & 0.723 \\
				\rowcolor[gray]{0.95}[\tabcolsep][\tabcolsep]
				\textbf{agony} & 0.528 & 0.518 & 0.565 & 0.523 & \sethlcolor{green}\hl{\textbf{0.498}} & 0.517 & 1.099 & 0.515 \\
				$\boldsymbol{\sigma_a}$ & 0.666 & 0.618 & 0.704 & \sethlcolor{green}\hl{\textbf{0.733}} & 0.669 & 0.687 & --   & 0.679 \\
				\rowcolor[gray]{0.95}[\tabcolsep][\tabcolsep]
				$\boldsymbol{\sigma_L}$ & -1.211 & -1.324 & -1.231 & -1.172 & -1.148 & \sethlcolor{green}\hl{\textbf{-1.109}} & --   & -1.111 \\
		\end{tabular}}
		\caption{\textbf{Results obtained from synthetic data in a static framework.} Performance comparison of the various models on a synthetic dataset where the ranks are fixed along time (static framework). The green highlighted values are the top performances for the considered metric. \Cref{tb:sem_static} contains the standard error of the above values. $\sigma_a$ and $\sigma_L$ cannot be applied to the W-L model, so there are no values for the metrics.}
		\label{tab:static_results}
	\end{table}
	
	\begin{table}[h!]
		\centering
		\begin{tabular}{c|cccccccc}
			\textbf{Metric} &     \textbf{Elo} &  \textbf{\nmdsr} &    \textbf{\mwsr} &     \textbf{\dsr} &      \textbf{SR} &      \textbf{TS} &       \textbf{W-L} &     \textbf{WHR} \\
			\hline
			\textbf{accuracy} &  0.0107 &  0.0109 &  0.0109 &  0.0109 &  0.0111 &  0.0101 &  0.0192 &  0.0099 \\
			\textbf{agony}    &  0.0243 &  0.0247 &  0.0250 &  0.0261 &  0.0259 &  0.0242 &  0.0552 &  0.0242 \\
			$\boldsymbol{\sigma_a}$  &  0.0059 &  0.0052 &  0.0099 &  0.0095 &  0.0085 &  0.0058 &        -- &  0.0056 \\
			$\boldsymbol{\sigma_L}$  &  0.0423 &  0.0049 &  0.0380 &  0.0399 &  0.0223 	&  0.0277 &        -- &  0.0251 \\
		\end{tabular}
		\caption{\textbf{Standard error of results from the synthetic data in a static framework.}}
		\label{tb:sem_static}
	\end{table}

	\section{Real Data Experiments}
	\paragraph*{Standard errors.} We report the standard errors of the experiments on the real datasets in \Cref{tb:sem_real}. These complement \Cref{tab:real_data_results} in the main manuscript.

	\begin{table}[htbp!]
		\centering
		\begin{tabular}{cc|cccccccc}
			\textbf{Dataset} & \textbf{Metric} &     \textbf{Elo} &  \textbf{\nmdsr} &    \textbf{\mwsr} &     \textbf{\dsr} &      \textbf{SR} &      \textbf{TS} &     \textbf{WHR} &       \textbf{W-L} \\
			\hline
			\multirow{4}*{\textbf{NBA}} & accuracy &  0.0048 &  0.0049 &  0.0046 &  0.0048 &  0.0047 &  0.0050 &  0.0047 &  0.0049 \\
			& agony &  0.0587 &  0.0588 &  0.0621 &  0.0577 &  0.0529 &  0.0576 &  0.0553 &  0.0665 \\
			& $\sigma_a$ &  0.0021 &  0.0026 &  0.0045 &  0.0045 &  0.0040 &  0.0024 &  0.0022 &        -- \\
			& $\sigma_L$ &  0.0154 &  0.0029 &  0.0075 &  0.0063 &  0.0041 &  0.0097 &  0.0082 &        -- \\
			\hline
			\multirow{4}*{\textbf{Chess}} & accuracy &  0.0213 &  0.0242 &  0.0246 &  0.0213 &  0.0214 &  0.0235&  0.0226 &  0.0119 \\
			& agony &  1.2323 &  1.8357&  1.6712 &  1.2301 &  1.2292 &  1.3637 &  1.3848 &  0.3262 \\
			& $\sigma_a$ &  0.0137 &  0.0134 &  0.0183 &  0.0179 &  0.0123 &  0.0144 &  0.0132 &        -- \\
			& $\sigma_L$ &  0.0773 &  0.0356 &  0.0672 &  0.0753 &  0.1090 &  0.0733 &  0.0526 &        -- \\
			\hline
			\multirow{4}*{\textbf{EPL}} & accuracy &  0.0061 &  0.0067 &  0.0073 &  0.0058 &  0.0063 &  0.0064 &  0.0062 &  0.0079 \\
			& agony &  0.0920 &  0.1292 &  0.1429 &  0.0906 &  0.0935 &  0.1172 &  0.1082 &  0.1424 \\
			& $\sigma_a$ &  0.0026 &  0.0035 &  0.0068 &  0.0059 &  0.0060 &  0.0030 &  0.0032 &        -- \\
			& $\sigma_L$ &  0.0249 &  0.0038 &  0.0139 &  0.0145 &  0.0112 &  0.0154 &  0.0141 &        -- \\
			\hline
			\multirow{4}*{\textbf{Serie A}}  & accuracy &  0.0048 &  0.0051 &  0.0051 &  0.0049 &  0.0050 &  0.0050 &  0.0047 &  0.0059 \\
			& agony &  0.0881 &  0.1227 &  0.1282 &  0.0856 &  0.0930 &  0.1252 &  0.1174 &  0.1617 \\
			& $\sigma_a$ &  0.0020 &  0.0013 &  0.0047 &  0.0044 &  0.0045 &  0.0024 &  0.0022 &        -- \\
			& $\sigma_L$ &  0.0176 &  0.0013 &  0.0102 &  0.0104 &  0.0060 &  0.0107 &  0.0093 &        -- \\
		\end{tabular}
		\caption{\textbf{Standard error of results from real data}}
		\label{tb:sem_real}
	\end{table}
	
	\section{Null Model Experiments}\label{sec:null_experiments}
	\paragraph*{Synthetic data.} We report results of the null model experiments where we permute the chronological order of synthetic dynamic data in  \Cref{fig:histogram_varying_noise} and of synthetic static data in \Cref{fig:histogram_static}.
	\paragraph*{Real data.} We report p-values on the null model experiments on real data in \Cref{tb:pvalues}.
	
	\begin{figure}[!htb]
		\centering
		{\includegraphics[width=0.85\linewidth,height=4cm]{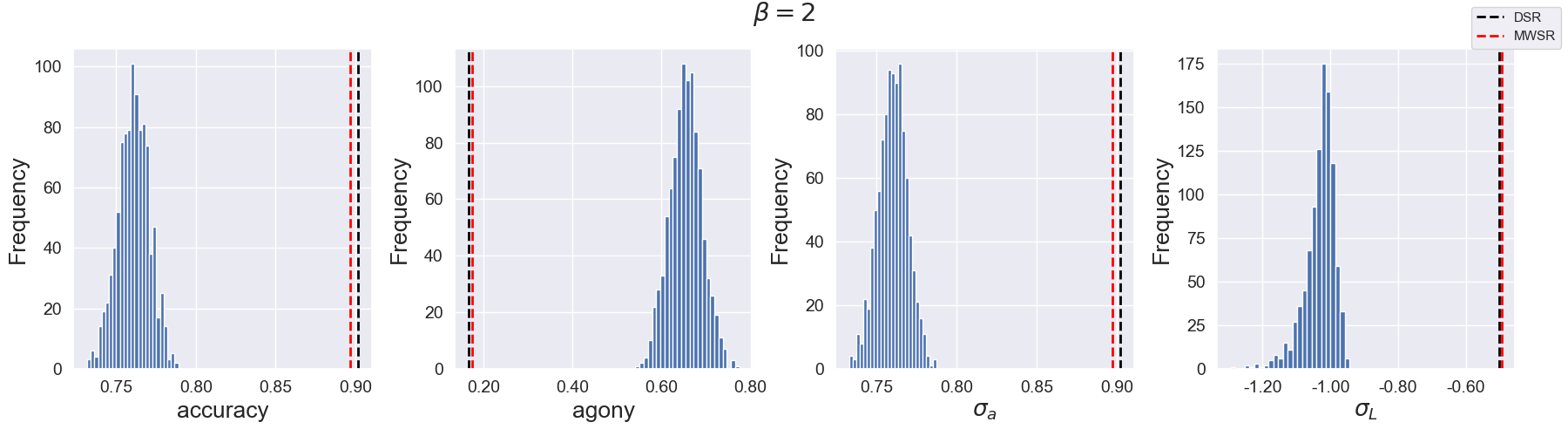}}\\
		{\includegraphics[width=0.85\linewidth,height=4cm]{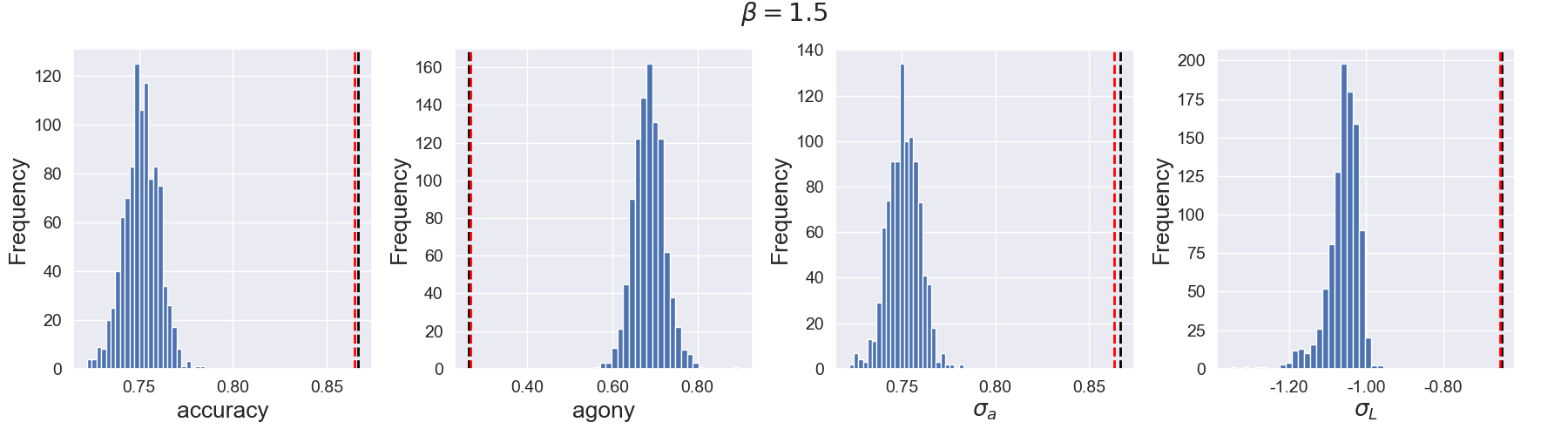}}\\
		{\includegraphics[width=0.85\linewidth,height=4cm]{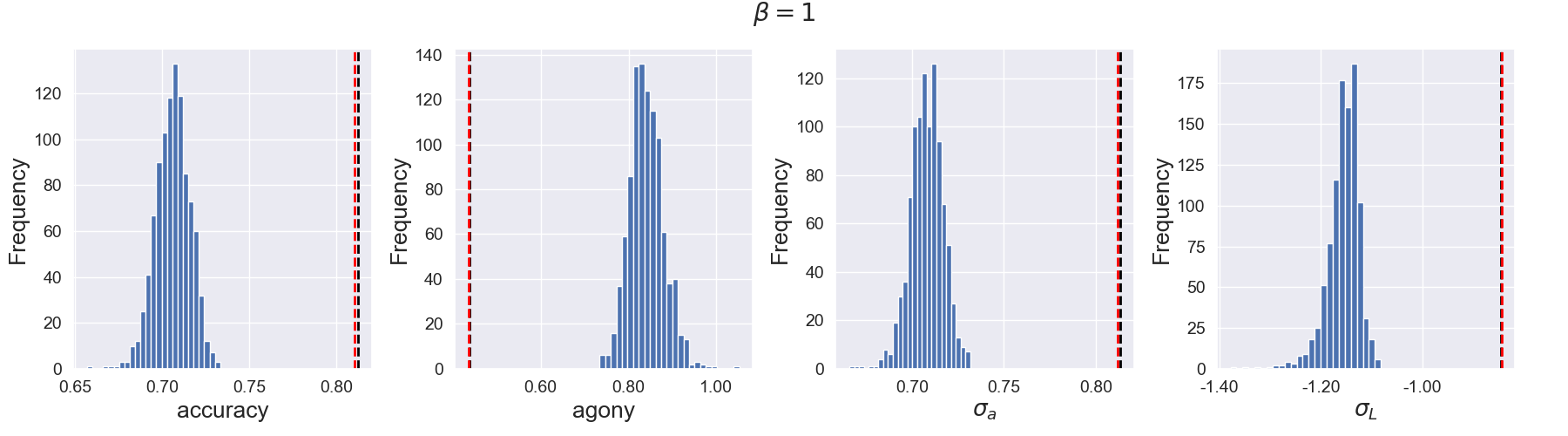}}\\
		{\includegraphics[width=0.85\linewidth,height=4cm]{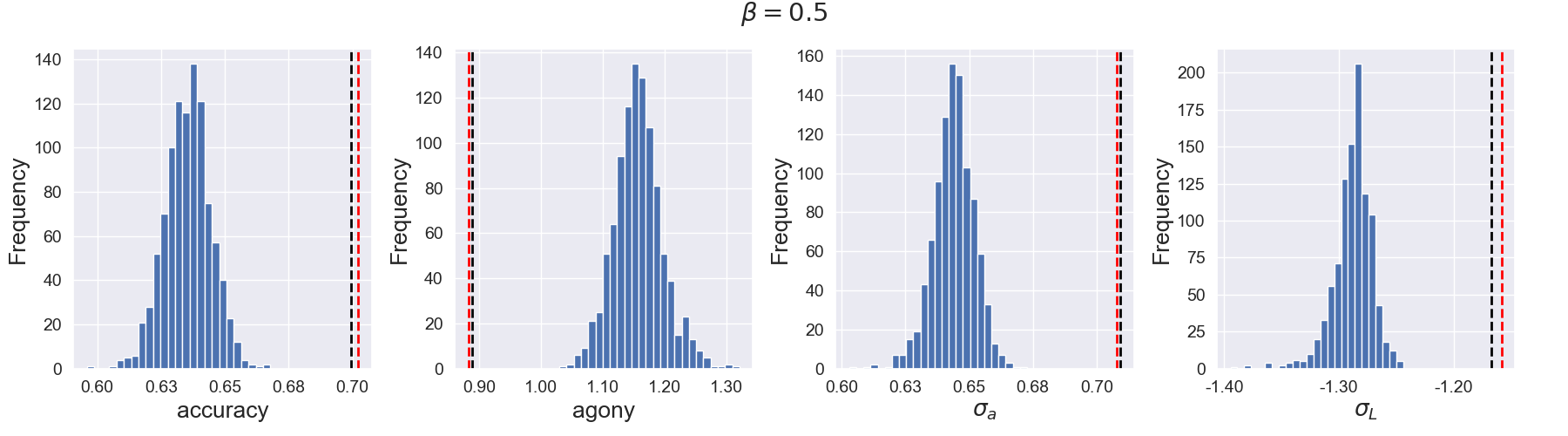}}\\
		{\includegraphics[width=0.85\linewidth,height=4cm]{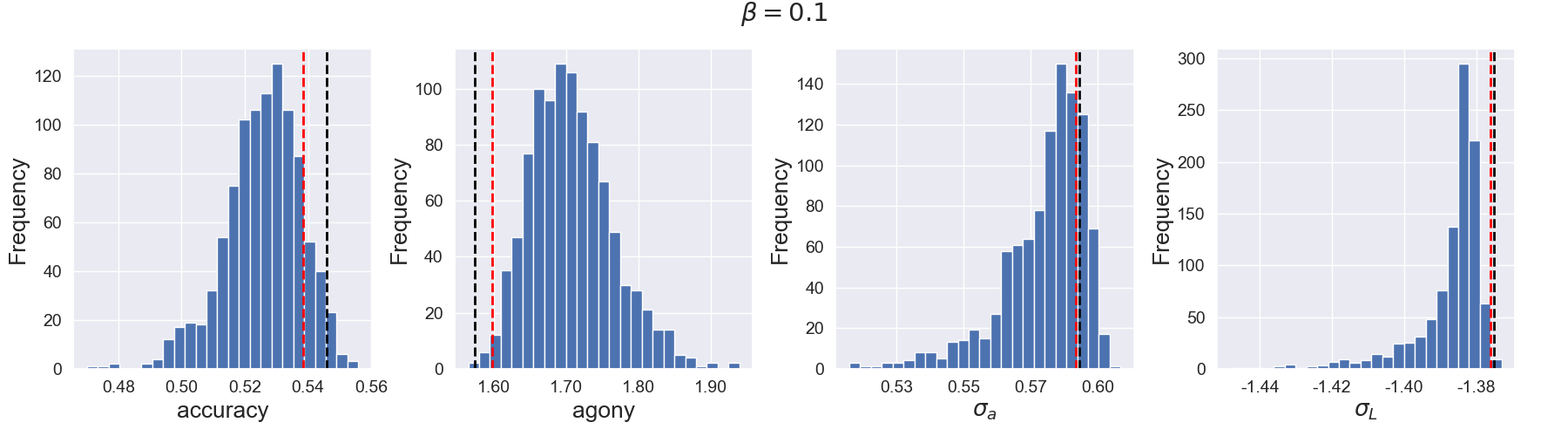}}
		\caption{\textbf{Null model results on the synthetic dataset with varying levels of density.} It is used to determine whether chronology is important. Each entry of the histogram is a different result of \dsr on the synthetic dataset where time-steps have been randomly permutated. 1000 permutations were considered. The black and red dotted lines represent the results of \dsr and \mwsr\ respectively on the regular, chronologically-ordered dataset.}
		\label{fig:histogram_varying_noise}
	\end{figure}
	
	\begin{figure}[!htb]
		\centering
		\includegraphics[width=\linewidth]{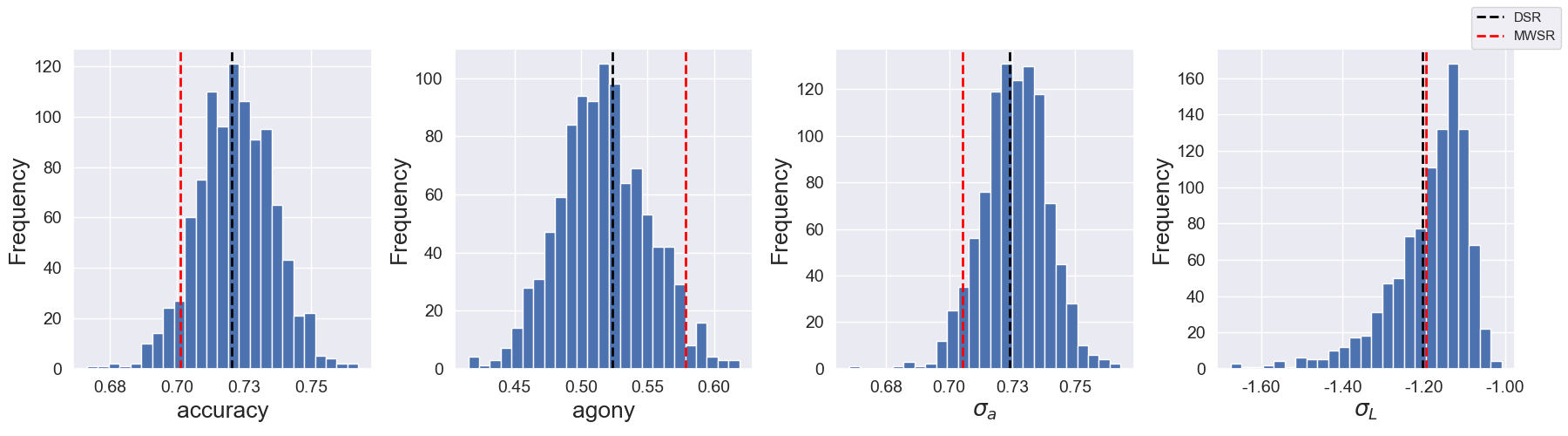}
		\caption{\textbf{Null model results on the synthetic dataset with static ranks.} It is used to determine whether chronology is important. Each entry of the histogram is a different result of \dsr on the synthetic dataset where time-steps have been randomly permutated. 1000 permutations were considered. The black and red dotted lines represent the results of \dsr and \mwsr\ respectively on the regular, chronologically-ordered dataset.}
		\label{fig:histogram_static}
	\end{figure}
	
	\begin{figure}[!htb]
		\centering
		{\includegraphics[width=0.9\linewidth,height=4.4cm]{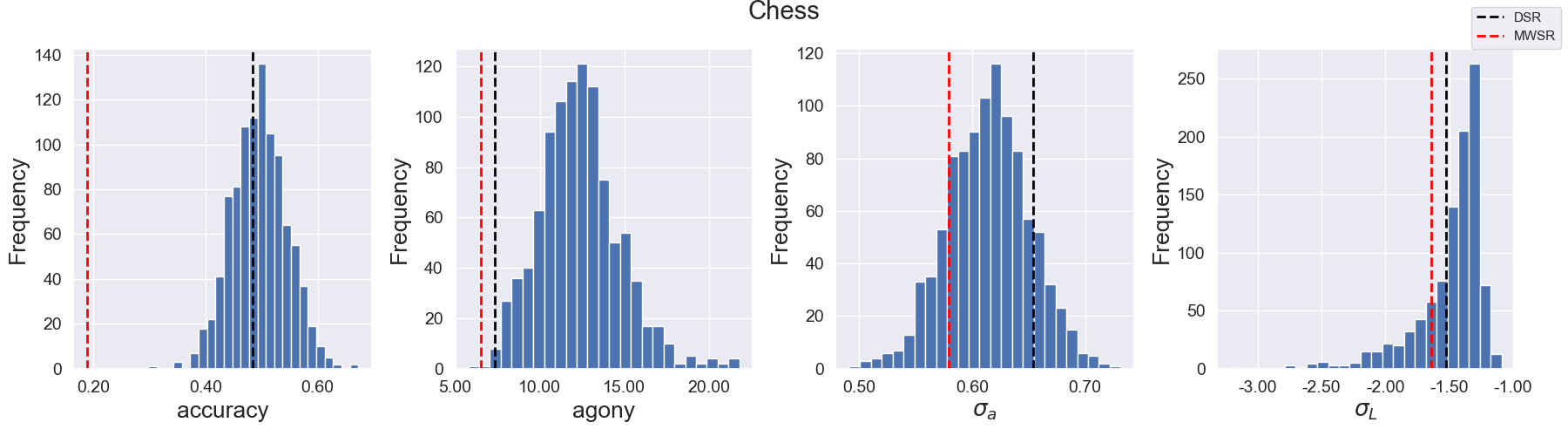}}\\
		{\includegraphics[width=0.9\linewidth,height=4.4cm]{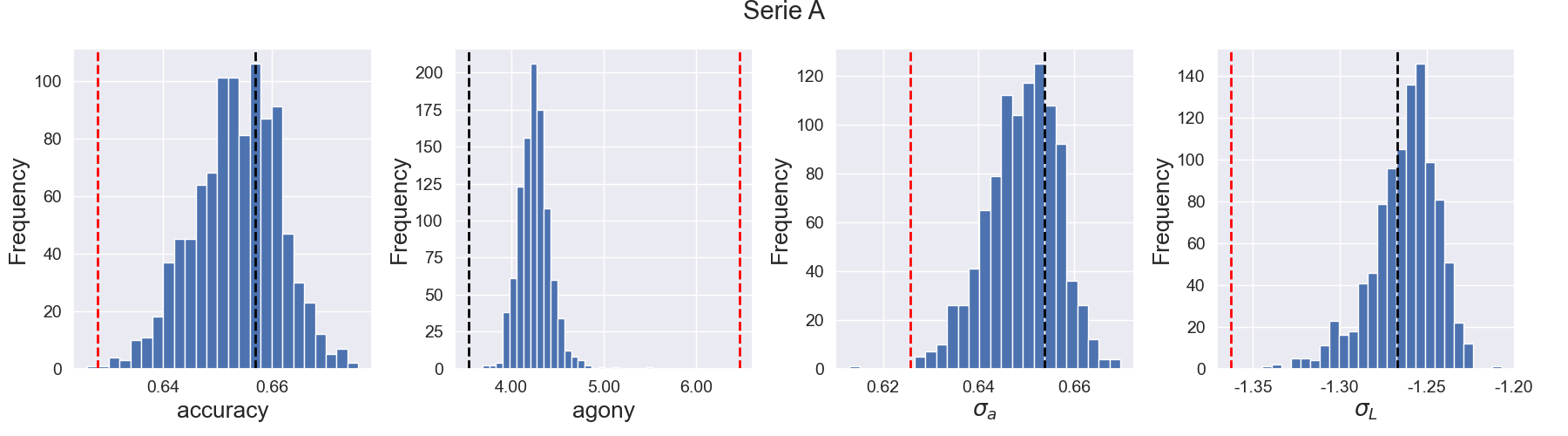}}\\
		{\includegraphics[width=0.9\linewidth,height=4.4cm]{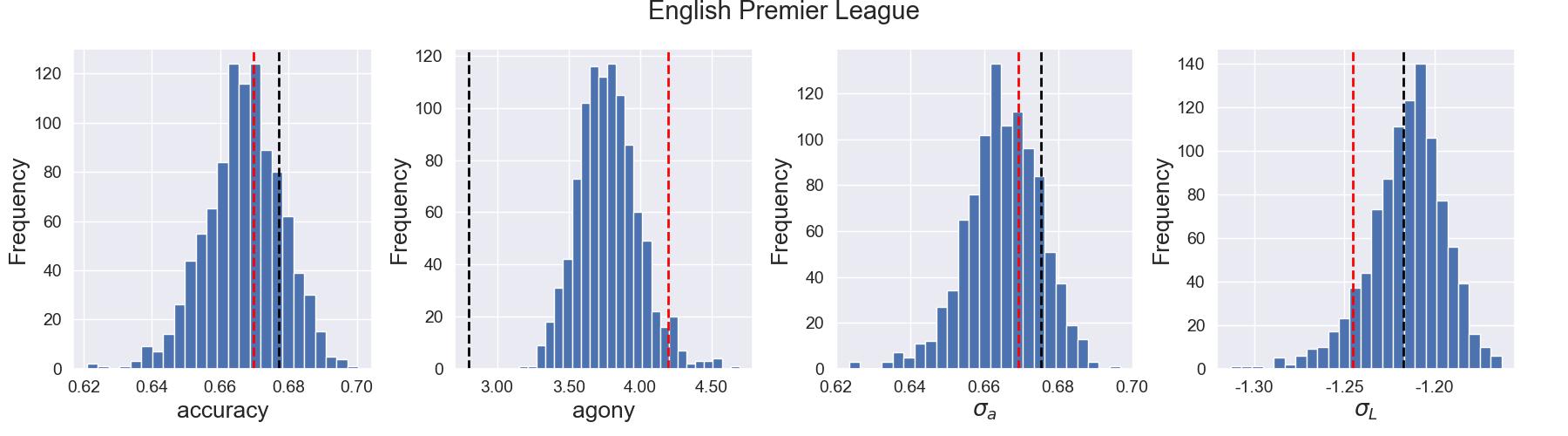}}
		\caption{\textbf{Null model results on the chess, EPL and Serie A datasets.} It is used to determine whether chronology is important. Each entry of the histogram is a different result of \dsr on the aforementioned datasets, where time-steps have been randomly permutated. 1000 permutations were considered. The black and red dotted lines represent the results of \dsr and \mwsr\ respectively on the regular, chronologically-ordered datasets.}
		\label{fig:histogram_real_extra}
	\end{figure}
	
	\begin{table}[h!]
		\centering
		{\renewcommand{\arraystretch}{1.2}
			\begin{tabular}{cc|cccc}
				Model &Metric	&   NBA & Chess  & EPL & Serie A\\
				\hline 
				\multirow{4}*{\dsr} &  accuracy & 0.0 & 0.594 & 0.183 & 0.359\\
				& agony &0.0&0.006&0.0&0.0\\
				&$\sigma_{a}$& 0.0&0.139 &0.155&0.282\\
				& $\sigma_{L}$ & 0.0 & 0.711 &0.578&0.644\\
				\hline
				\multirow{4}*{\mwsr} &  accuracy & 0.0 & 1.0 &0.411 & 0.999\\
				& agony &0.0&0.001&0.959&1.0\\
				&$\sigma_{a}$& 0.0&0.832 &0.340&0.999\\
				& $\sigma_{L}$ & 0.0 & 0.797 &0.911&1.0\\
		\end{tabular}}
		\caption{\textbf{Null model p-value results on real data.} Illustrated are the number of times (as a percentage) that the metric value on the randomized dataset is better than on the chronologically-ordered dataset. Results are calculated over $1000$ permutations.}
		\label{tb:pvalues}
	\end{table}
	
\section*{Acknowledgements}
We thank Jean-Gabriel Young for his feedback on the boundary conditions of the Offline Dynamical SpringRank model.
\end{widetext}

\end{document}